\documentclass[12pt]{iopart}
\usepackage{iopams}
\usepackage{graphicx}
\usepackage[T1]{fontenc}
\usepackage{ae}
\usepackage[latin1]{inputenc}
\usepackage{color}

\newcommand{\Eins}
           {\;\smash{\raisebox{-0.5ex}{$\!\!\stackrel{\!\mbox{1}
            \hspace{-0.4ex}\rule[0.0ex]{0.06ex}{1.60ex}}{ }$}}}
\newcommand{\op}[1]{%
    \fontdimen12\textfont3=2pt\fontdimen12\scriptfont3=1.4pt%
    \!\null\mathop{\vphantom{#1}\smash{#1}}\limits_{\sim}\null\!}
\newtheorem{lemma}{Lemma}

\newtheorem{prop}{Proposition}

\newtheorem{theorem}{Theorem}
\newtheorem{cor}{Corollary}

\begin{document}

\title[Coupled trimers]
{Exact ground states for coupled spin trimers}

\author{Heinz-J\"urgen Schmidt$^1$
 \footnote[3]{Correspondence should be addressed to
hschmidt@uos.de} and Johannes Richter$^2$ }

\address{$^1$ Universit\"at Osnabr\"uck, Fachbereich Physik,
Barbarastr. 7, D - 49069 Osnabr\"uck, Germany\\
$^2$Institut f\"ur Theoretische Physik, Otto-von-Guericke-Universit\"at Magdeburg,\\
PF 4120, D - 39016 Magdeburg, Germany }

\begin{abstract}
We consider a class of geometrically frustrated Heisenberg spin systems which admit
exact ground states. The systems consist of suitably coupled antiferromagnetic spin trimers with
integer spin quantum numbers $s$ and their ground state $\Phi$ will be the product state of
the local singlet ground states of the trimers. We provide linear equations for the inter-trimer
coupling constants which are equivalent to $\Phi$ being an eigenstate of the corresponding
Heisenberg Hamiltonian and sufficient conditions for $\Phi$ being a ground state. The classical case
$s\to\infty$ can be completely analyzed. For the quantum case we consider a couple of examples, where
the critical values of the inter-trimer couplings are numerically determined. These examples include
chains of corner sharing tetrahedra as well as certain spin tubes. $\Phi$ is proven to be gapped in the
case of trimer chains. This follows from a more general theorem on quantum chains with product ground states.
\end{abstract}

\pacs{75.10.b, 75.10.Jm}


\maketitle

\section{Introduction}
\label{sec1}

The effects of frustration in Heisenberg magnets have attracted much
interest over the past few decades \cite{rev1,rev2}. Geometrically
frustrated quantum antiferromagnets (AF) are an excellent
play-ground for studying novel quantum many-body phenomena. We
mention quantum spin-liquid phases, valence-bond crystal phases,
order-by-disorder
phenomena, lattice instabilities to name just a few.\\

Moreover, in recent years there has been remarkable progress in
synthesizing magnetic materials \cite{lemmens}. Even exotic
structures such as the star \cite{star} or the maple-leaf lattices
\cite{maple} have been synthesized. Hence, the investigation of
exotic lattice structures being on the
first glance purely academic might become relevant for experimental studies.\\

The theoretical investigation of frustration effects in quantum spin
anti-ferromagnets usually meets new difficulties; e.~g.~the quantum Monte
Carlo method suffers from the sign problem for frustrated systems.
Exact statements for interacting quantum-many body systems are rare and, therefore,
new rigorous results are of considerable interest to improve their
understanding.
Moreover, solvable models may serve as test grounds for approximate methods.\\

Starting with the seminal papers of
Majumdar-Ghosh \cite{MG69}
and Shastry-Sutherland \cite{Shastry81},
a famous class of spin systems with exact ground states has thoroughly been investigated,
see, for example, \cite{DMS}. These systems consist of
$N$ suitably coupled AF dimers such that the product state of the local $S=0$ dimer ground states
remains the ground state of the total system. For this it is necessary that the coupling constants
between the dimers satisfy certain linear equations \cite{Schmidt05}. A generalization to chains with
trimerized ground states using $SU(3)$ spins and bilinear-biquadratic Hamiltonians has recently been
published \cite{Greiter}.
Another approach using matrix product states as exact ground states \cite{ Zittartz} has also been mainly
applied to anisotropic spin systems.
\\

In this paper we will rather consider isotropic (Heisenberg) spin systems and
extend the basic idea underlying the  Majumdar-Ghosh-Shastry-Sutherland ground states to systems with
AF trimers as building blocks. These systems will be geometrically frustrated.
If the individual spin quantum number $s$ is integer,
the ground state of a uniform (or almost uniform) AF trimer will be again a non-degenerate
singlet state, i.~e.~having $S=0$. The product state $\Phi$ of these local ground states
will be an eigenstate of the Heisenberg Hamiltonian of $N$ trimers if and only if the
coupling constants satisfy certain linear constraints completely analogous to the dimer case.
Moreover, if the coupling between the dimers is not too strong, $\Phi$ will be the ground state
of the total system, called the {\it trimerized ground state} (TGS).
Of course, the question arises what precisely is meant by "not too strong"?
In general,
the precise domain of systems admitting TGS can only be numerically
investigated for given examples and lattice structures.
In a sense, we thus invert the usual strategy to numerically find ground states for given spin systems.
We define a certain state $\Phi$ and numerically calculate the
coupling constants for systems which have $\Phi$ as their ground state.
However, we have also derived some general rigorous statements on systems with
trimerized ground states.
These statements mainly concern trimerized ground states in the classical limit $s\to\infty$, which can be completely
understood, and some sufficient conditions for trimerized ground states in the quantum case.
Moreover, we prove that in the case of trimer chains with trimerized ground state,
$\Phi$ will be gapped for an extended domain of coupling constants.
We stress that our investigation is restricted to the case of integer $s$, since for $s$ being half integer
the ground state of the AF trimer will be degenerate, see, e.~g.~\cite{caspers}.
We mention that trimerized states of the kind explained above have also been considered as
approximate ground states of certain modified 2D kagome lattices \cite{Cai09}.
\\

The paper is organized as follows. In section \ref{secS} we
summarize the basic definitions and main results. Section
\ref{secE} is devoted to a couple of examples, starting from a
single trimer, followed by a pair of trimers and then passing to
$1$-dimensional chains of trimers. The latter examples cover
systems like chains of corner-sharing tetrahedra \cite{tetrahedra}
and spin tubes \cite{tubes}, which currently attract a lot of
attention in the literature, albeit, as a rule, not under the
aspect of exact ground states. The study of the examples led us to
some conjectures about the domain of the coupling constants for
systems admitting trimerized ground states and its dependence on
$s$ and $N$. In order to clearly distinguish these conjectures
from the mentioned rigorous statements we have presented the
latter in a separate section \ref{sec2}. We decided to give
detailed proofs of these statements only in those cases where they
markedly differ from the analogous proofs for the dimer case in
\cite{Schmidt05}.





\section{Basic definitions and summary of main results}\label{secS}


We consider systems of $3N$ spins with
one and the same individual integer spin
quantum number $s=1,2,3,\ldots$ which are grouped into $N$ fixed triples
(``trimers"). To indicate this grouping the spins will be denoted
by indices $\mu =(i,\delta)$ where $i=1,\ldots, N$ is the trimer
index and $\delta = 0,1,2$ distinguishes between the three spins
belonging to the same trimer. Further we consider Heisenberg
Hamiltonians
\begin{eqnarray} \label{S1a}
\op{H}({\mathbb J})&=& \sum_{\mu\nu}J_{\mu\nu}\op{\bi{s}}_\mu\cdot \op{\bi{s}}_\nu\\
&=&
\sum_{i,j}\sum_{\delta, \epsilon}J_{i\delta,\,j\epsilon}\;\op{\bi{s}}_{i\delta}\cdot \op{\bi{s}}_{j\epsilon}
\;,
\end{eqnarray}
where $\op{\bi{s}}_\mu=(\op{s}_\mu^{(1)},\op{s}_\mu^{(2)},\op{s}_\mu^{(3)})$ denotes
the $\mu$-th spin observable and ${\mathbb J}$ the $3N\times
3N$-matrix of real exchange parameters or coupling constants
$J_{\mu\nu}$ satifying
\begin{equation} \label{S2}
J_{\mu\mu}=0,\quad J_{\mu\nu}=J_{\nu\mu}\mbox{ for all } \mu,\nu=1,\ldots,3N
\;.
\end{equation}
All operators act on a $(2s+1)^{3N}$-dimensional Hilbert space
${\mathcal H}=\bigotimes_{\mu=1}^{3N}{\mathcal H}_\mu$.
If the spin quantum number $s$ is fixed, we may
identify a spin system with its matrix ${\mathbb J}$.
Note that due to the matrix notation each scalar product
$\op{\bi{s}}_\mu\cdot \op{\bi{s}}_\nu$ occurs twice in the
Hamiltonian (\ref{S1a}). In order to comply
with the usual notation we have therefore introduced the factor
$\frac{1}{2}$ in some examples of section \ref{secE}.\\

For any trimer with index $i$ let $[i0,i1,i2]$ denote the ground state
of the AF trimer
$\op{H}_0=\lambda_i\sum_{\delta,\epsilon}\op{\bi{s}}_{i\delta}\cdot\op{\bi{s}}_{i\epsilon},\;\lambda_i>0$
which is unique up to a phase factor. If the trimer index $i$ is irrelevant, it will be simply
denoted by $[0,1,2]$. For general $s$ this state can be written in terms of the Wigner-$3j$-symbol as
\begin{eqnarray}
[0,1,2]&=& \sum_{m_0,m_1=-s}^s
\left(
\begin{array}{ccc}
s&s&s\\
m_0 & m_1 & -m_0-m_1
\end{array}
\right)
|m_0,m_1,-m_0-m_1\rangle
\;,
\end{eqnarray}
using the eigenbasis $|m\rangle,\; m=-s,\ldots,s$ of
$\op{\bi{s}}_\mu^{(3)}$ and the corresponding product bases.\\
The state $[i0,i1,i2]$ will remain the ground state of the trimer $i$
even when its Hamiltonian is suitably disturbed, see section \ref{secE1}.
Let $\mathcal{C}_i^s$ denote the set of all $(J_{i0},J_{i1},J_{i2})$ where
this is the case.\\

The ground state of a system of $N$ unconnected AF trimers
satisfying $(J_{i0},J_{i1},J_{i2})\in\mathcal{C}_i^s$ for $i=1,\ldots,N$
is the product state
\begin{equation}\label{S4}
\Phi^s \equiv \bigotimes_{i=1}^N [i0,i1,i2] \;,
\end{equation}
called the {\it trimerized state}; it has the total spin quantum
number $S=0$. A system ${\mathbb J}$ is said to admit trimerized
ground states (TGS), or to have the TGS property, iff $\Phi^s$ is
a ground state of $\op{H}({\mathbb J})$, i.~e.~iff
\begin{equation}\label{S5}
\langle \Phi^s | \op{H}({\mathbb J}) \Phi^s\rangle
\le
\langle \Psi |\op{H}({\mathbb J}) \Psi\rangle
\end{equation}
for all $\Psi\in{\mathcal H}$ with $||\Psi||=1$.
Let ${\mathcal C}_\Phi^s$ denote the set of all spin systems ${\mathbb J}$ with the
TGS property. If the quantum number $s$ is understood,
we suppress it and write simply $\Phi$ and ${\mathcal C}_\Phi$. \\

For ${\mathbb J}\in {\mathcal C}_\Phi$ it is necessary that $\Phi$
will be an eigenstate of $\op{H}({\mathbb J})$. This turns out to be true
if and only if the inter-trimer coupling constants fulfil the
relations
\begin{eqnarray}\label{S6a}
J_{i0,j0}+ J_{i1,j1}&=&J_{i0,j1}+J_{i1,j0}\\ \label{S6b}
J_{i0,j0}+ J_{i1,j2}&=&J_{i0,j2}+J_{i1,j0}\\ \label{S6c}
J_{i0,j0}+ J_{i2,j1}&=&J_{i0,j1}+J_{i2,j0}\\ \label{S6d}
J_{i0,j0}+ J_{i2,j2}&=&J_{i0,j2}+J_{i2,j0}
\end{eqnarray}
for all $1\le i < j \le N$.
The corresponding eigenvalue
\begin{equation}\label{S7}
E=-s(s+1)\left(\sum_{i=1}^N J_{i0,i1}+
J_{i0,i2}+J_{i1,i2}\right).
\end{equation}
is independent of the inter-trimer coupling.
Since  (\ref{S6a})-(\ref{S6d}) is a system of four independent linear
equations, the set of all
real, symmetric $3N\times 3N$-matrices satisfying (\ref{S6a})-(\ref{S6d}) and
$J_{\mu\mu}=0$ for all $\mu=1,\ldots,3N$ will be a linear space of dimension
$3N+5 {N\choose 2}=\frac{N}{2}(5N+$1), denoted by $\mathcal{J}_\Phi$.
The set $\mathcal{C}_\Phi$ of TGS systems will form a convex cone
embedded in the linear space $\mathcal{J}_\Phi$, since the condition (\ref{S5})
is invariant under positive linear combinations of $\mathbb{J}$'s , see also \cite{Schmidt05}.
At the boundary of $\mathcal{C}_\Phi$, the trimerized ground state $\Phi$ will become
degenerate, i.~e.~there will exist ``competing"
ground states which have a lower energy than
(\ref{S7}) if $\mathbb{J}$ crosses the boundary of $\mathcal{C}_\Phi$.
Sometimes we will also use the symbol $\stackrel{\circ}{\mathcal{C}}_\Phi$ in order
to denote the open convex cone of spin systems $\mathbb{J}$ where $\Phi$ is a non-degenerate ground state.
\\

The conditions (\ref{S6a})-(\ref{S6d}) still include interesting spin structures
such as corner-sharing tetrahedra and spin tubes to be considered in section \ref{secE3}.
In some of these chains additional symmetries arise which allow the description
by an equivalent ladder model of composite spins, see \ref{secE31} and \ref{secE32}.
However, the TGS property is independent of this additional symmetry as shown
by the example of a certain spin tube in section \ref{secE33}.
\\

It is clear that the choice of dimensionless numbers for the $J_{\mu\nu}$ in the
examples implies the introduction of appropriate units for energy and temperature
in order to apply the Heisenberg model to real systems. In this sense, temperature
becomes a dimensionless quantity in the thermodynamic calculations of section \ref{secE2}
which illustrate some physical consequences of the existence of TGS ground states.
In most cases the domain of TGS systems, i.~e.~the shape of the cone
$\mathcal{C}_\Phi^s$ can only be determined numerically. Exceptions
are the single trimer case where $\mathcal{C}_\Phi^s$ can be calculated analytically,
see section \ref{secE1}, and the two-trimer case with $s=1$, see section \ref{secE2}.
In all
examples which are considered in section \ref{secE} there is some
evidence that these cones shrink with increasing $s$. Note that in the
classical limit $\mathcal{C}_\Phi^\infty$ will be degenerate, since the necessary conditions
for the $J_{\mu\nu}$ are stronger in this case, see section \ref{sec2.2}.
In contrast, for the chains of trimers considered in section
\ref{secE3}, the dependence of $\mathcal{C}_\Phi^s$ on $N$ is only weak.
Moreover, we found that $\mathcal{C}_\Phi^s$ is slightly expanding if $N$ grows.
This is an indication that the limit of
$\mathcal{C}_\Phi^s$ for $N\to\infty$ is ``non-degenerate", in the
sense that it does not converge to a lower-dimensional domain and that the trimerized ground state is gapped.
We will rigorously prove these properties in section \ref{sec2.4}. Our class of models thus supports Haldane's
conjecture \cite{Hal} that integer spin chains possess a unique ground state and
an energy gap between the ground state and the excited states. The famous AKLT model \cite{AKLT}
is also an $s=1$ spin chain with a unique gapped ground state but its anti-ferromagnetic
Heisenberg Hamiltonian is modified by a biquadratic term. For the $s=1/2$ Heisenberg spin chain
the absence of an energy gap has been proven in the `Lieb-Schultz-Mattis theorem' \cite{LSM}.
\\

In the classical case $s\to\infty$ an analogous definition of TGS systems is possible, see section \ref{sec2}.
In this case $\Phi$ consists of all spin configurations with a mutual angle of $120^\circ$ between spin
vectors of the same trimer.
For classical TGS systems the inter-trimer coupling must be necessarily uniform
and hence can be described by a symmetric $N\times N-$ matrix $\mathbb{G}$.
We have the result that a classical system has the TGS property if and only if $\mathbb{G}$ is positive
semi-definite. Thus the classical case is completely understood.

\section{Examples}
\label{secE}

For readers less interested in the mathematical details of our rigorous
analysis that will be presented  in more detail below in
section \ref{sec2} we first present some examples.
In particular, the general statements listed above can be used to discuss
certain specific chain-like models, such as  chains
of corner sharing tetrahedra as well as various spin tubes. We begin with
some more elementary examples.

\subsection{One Trimer}
\label{secE1}
\begin{center}
\begin{figure}
  \includegraphics[width=13cm]{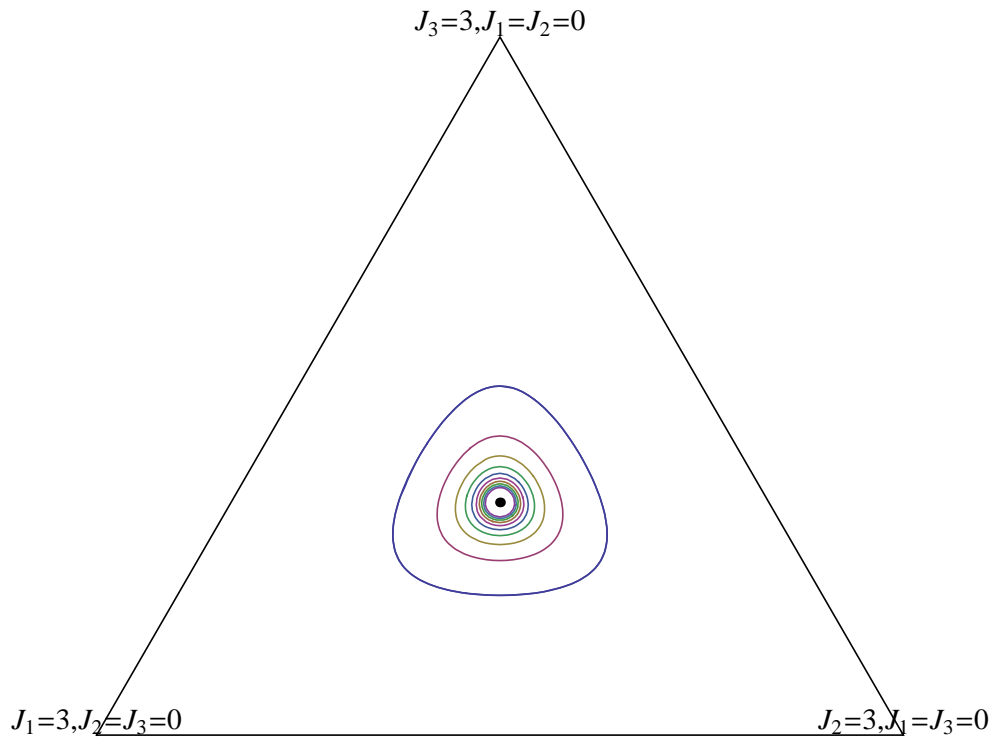}
\caption{\label{fige1}Representation of the neighborhood of
$J_1=J_2=J_3=1$ in which the state $[1,2,3]$ remains the ground
state of (\ref{E1.1}) where $s=1,\ldots,10$ starting with the
outermost curve. The values of $J_1,J_2,J_3$ are restricted to the
plane $J_1+J_2+J_3=3$. The curves have been calculated according to
(\ref{E1.2}) and have been confirmed numerically for $s=1,\ldots,5$.
The point at the center of the figure with coordinates
$(1,1,1)$ corresponds to the classical
limit.}
\end{figure}
\end{center}

We consider three spins with integer spin quantum number $s$ and Heisenberg Hamiltonian
\begin{equation}\label{E1.1}
\op{H}=
\sum_{\mu,\nu=1}^3J_{\mu\nu}\op{\bi{s}}_\mu\cdot \op{\bi{s}}_\nu
=J_1 \op{\bi{s}}_2\cdot \op{\bi{s}}_3+
J_2 \op{\bi{s}}_3\cdot \op{\bi{s}}_1+J_3 \op{\bi{s}}_1\cdot \op{\bi{s}}_2
\;,
\end{equation}
where we have relabeled the coupling constants in order to keep the following formulas readable.
Let $[1,2,3]$ denote the unique state with vanishing total spin, $S=0$.
It is an eigenstate of (\ref{E1.1}),
since the eigenspaces
of $\op{S}^2$ are invariant under $\op{H}$. For certain values of $J_1,J_2,J_3$, $[1,2,3]$ is even the ground state of
$\op{H}$, e.~g.~for $J_1=J_2=J_3=1$. These values of $J_1,J_2,J_3$ form an closed convex cone
$\mathcal{C}^s$ in the $3$-dimensional  $(J_1,J_2,J_3)$-space. At the boundary of $\mathcal{C}^s$ the ground state of $\op{H}$ becomes degenerate.
Recall that $\stackrel{\circ}{\mathcal{C}^s}$ denotes the open subset of $\mathcal{C}^s$ where $[1,2,3]$ is the non-degenerate ground state.
The form of the
cone can be calculated using computer-algebraic software and the well-established assumption
that the competing state has the quantum number $S=1$. It is given by the following inequalities
\begin{eqnarray}\nonumber
(J_1,J_2,J_3)\,\in\,\stackrel{\circ}{\mathcal{C}^s} &\Leftrightarrow& \\ \nonumber
 \frac{1}{2}s(s+1)(J_1+J_2+J_3)(J_1J_2+J_2J_3+J_3J_1)&<&
(1+\frac{9}{2}s(s+1)) J_1J_2J_3 \\ \label{E1.2}
\mbox{ and } J_1,J_2,J_3> 0 \;.&&
\end{eqnarray}
The intersection of $\mathcal{C}^s,\; s=1,\ldots,10$ with the plane $J_1+J_2+J_3=3$
is depicted in figure \ref{fige1}.
By Taylor expansion of (\ref{E1.2}) one can show that, for increasing values of $s$,
the cones $\mathcal{C}^s$ approach a circular
form centered at the half line $J_1=J_2=J_3>0$ with a cone angle of
\begin{equation}
\vartheta=\arctan\sqrt{\frac{2}{3(1+3s(s+1))}}
\;.
\end{equation}
Hence the cones $\mathcal{C}^s$ will shrink
and approach their classical limit $J_1=J_2=J_3>0$ for $s\rightarrow\infty$,
see section \ref{sec2}.

\subsection{Two Trimers}
\label{secE2}

\begin{center}
\begin{figure}
\begin{center}
\includegraphics[width=8cm]{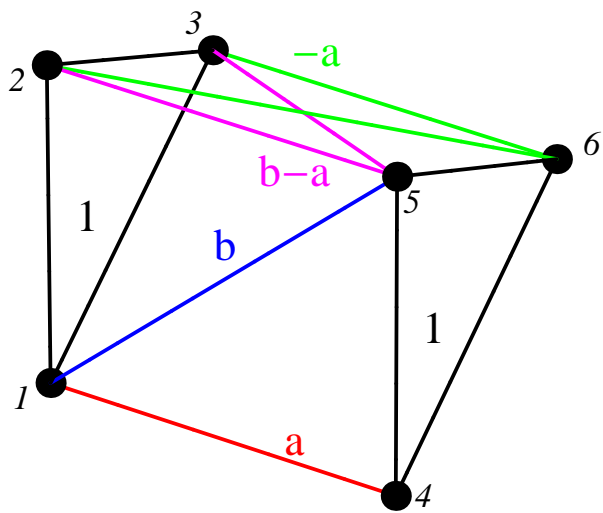}
\end{center}
\caption{\label{fige2}Two coupled trimers satisfying (\ref{E2.2a}),(\ref{E2.2b}).
The intra-trimer coupling (black lines) is set to one. The inter
trimer coupling strengths are $a$ (red line), $-a$ (green line), $b$ (blue line) and
$b-a$ (magenta line).}
\end{figure}
\end{center}

Next we consider two trimers, i.~e.~six spins grouped into two triples with indices $(1,2,3)$ and $(4,5,6)$
and Heisenberg Hamiltonian
\begin{equation}\label{E2.1}
\op{H}=\sum_{\mu,\nu=1}^6 J_{\mu\nu} \op{\bi{s}}_\mu \cdot \op{\bi{s}}_{\nu}
\;.
\end{equation}
We ask whether $\Phi\equiv [1,2,3]\otimes[4,5,6]$ will be a ground
state of $\op{H}$. $\Phi$ is then called the {\em trimerized ground
state}. First, we note that $\Phi$ need not be an eigenstate of
$\op{H}$ unless the $J_{\mu\nu}$ do not satisfy equations
(\ref{S6a})-(\ref{S6d}). These equations
will be derived in section \ref{sec2}. They can be expressed in the
following way: Let $(X,Y)$ be a pair of spins belonging to the first trimer
and $(x,y)$ another pair belonging to the second one.
Then the two
sums $J_{Xx}+J_{Yy}$ and $J_{Xy}+J_{Yx}$ must be equal. This is a
kind of balance condition completely analogous to the corresponding
condition in the case of two dimers, see \cite{Schmidt05}. It is
satisfied for all combinations of spin pairs if and only if $\Phi$
is an eigenstate of $H$. Actually, only four conditions of the above
form have to be postulated, since the other will follow then, see
section \ref{sec2}. The coupling constants within the same trimer
are not constrained, hence we are left with a $3+3+5=11$-dimensional
linear space of independent
coupling constants $J_{\mu\nu}$, which will be called $\mathcal{J}$. It is independent of $s$.\\
Again, the linear space $\mathcal{J}$ will contain a closed convex cone $\mathcal{C}^s$ of values $J_{\mu\nu}$ such that $\Phi$
will be a ground state of the corresponding Hamiltonian (\ref{E2.1}). The example of two unconnected trimers
with the respective non-degenerate ground states $[1,2,3]$ and $[4,5,6]$ shows that $\mathcal{C}^s$ is not empty.
But it seems hopeless
to analytically calculate $\mathcal{C}^s$ except for  $s=1$. Thus we will here present some mixture of numerical,
computer-algebraic  and semi-analytical results.
Neglecting a positive overall factor in (\ref{E2.1}) we still have a $10$-dimensional manifold of possible
$J_{\mu\nu}$-values which is difficult to visualize. We will hence
confine ourselves to some two-dimensional subspace of $\mathcal{J}$
defined by (see also figure \ref{fige2})
\begin{eqnarray}\label{E2.2a}
J_{12}&=&J_{13}=J_{23}=J_{45}=J_{46}=J_{56}=\frac{1}{2}
\mbox{ and }\\ \label{E2.2b}
J_{14}&=&-J_{26}=-J_{36}=\frac{a}{2},\;
J_{15}=\frac{b}{2},\;J_{25}=J_{35}=\frac{b-a}{2}
\;,
\end{eqnarray}
where obviously the second equation fulfils the general conditions
(\ref{S6a})-(\ref{S6d}). Note that the
coupling strength between spins indicated in figure \ref{fige2}
equals twice the values of the $J_{\mu\nu}$ due to our definition of
the Hamiltonian (\ref{E2.1}).

\begin{center}
\begin{figure}
\begin{center}
  \includegraphics[width=10cm]{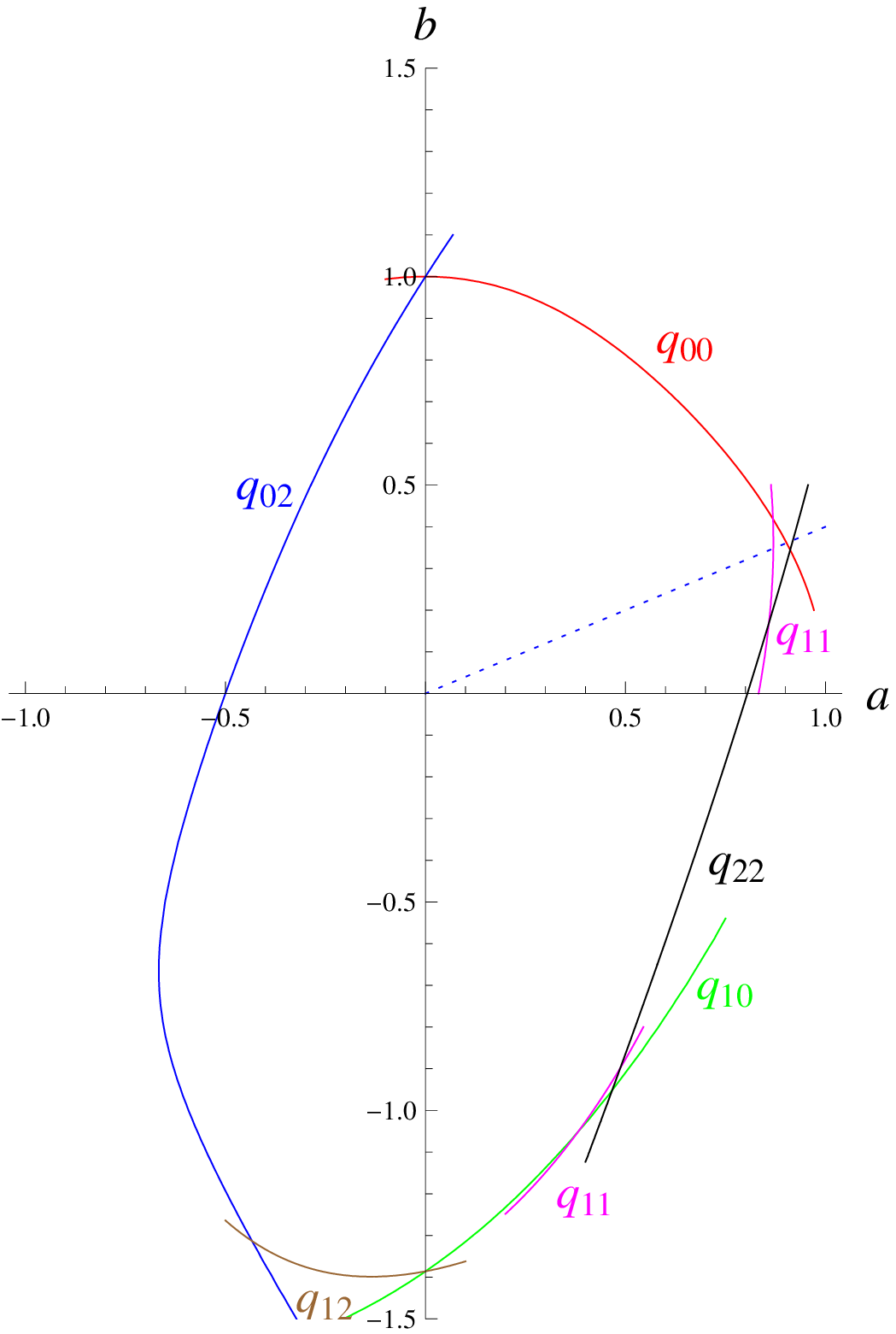}
\end{center}
\caption{\label{fige3a}Representation of the set $\mathcal{T}^s$, $s=1$,
of points with coordinates $(a,b)$ such that the singlet product state
$\Phi$ will be a ground state of the Hamiltonian (\ref{E2.1})
corresponding to (\ref{E2.2a},\ref{E2.2b}).
The boundary of $\mathcal{T}^1$ consists of pieces of $6$ intersecting smooth curves $q_{ij}$ which have been
calculated by computer-algebraic means. Note that $q_{11}$ contributes twice to the boundary.
The indices $i,j$ of $q_{ij}$ refer to the quantum
numbers $S\equiv i$ and $S_{23}\equiv j$ of the corresponding competing states. The dotted line
defined by $b=0.4\,a$ will be used for thermodynamic calculations, see figure \ref{fige3c}. }
\end{figure}
\end{center}
The set $\mathcal{T}^s$ of points with coordinates $(a,b)$ such that $\Phi$
will be a ground state of the corresponding Hamiltonian is again
a convex set. It is represented in figure \ref{fige3b} for the values $s=1,2,3,4,5$.
The case $s=1$ is just within the practical limits of computer-algebraic methods. The boundary of
$\mathcal{T}^1$ consists of pieces of $6$ intersecting curves given by equations of the form
$q_{ij}(a,b)=0$  where the $q_{ij}$ are polynomials in the variables $a,b$ with integer coefficients.
The three simplest cases are
\begin{equation}
q_{00}= 4 - 4 a - a^2 + a^3 - 4 b + 5 a b - a^2 b - b^2 - a b^2 + b^3
\,,
\end{equation}
\begin{equation}
q_{02}= 8 + 28 a + 28 a^2 + 8 a^3 - 8 b - 14 a b - 5 a^2 b - 2 b^2 -
5 a b^2 + 2 b^3
\;,
\end{equation}
and
\begin{eqnarray}\nonumber
q_{10}&=& 128 - 128 a - 74 a^2 + 85 a^3 - 2 a^4 - 11 a^5 + 2 a^6 -
  128 b + 226 a b- 29 a^2 b\\ \nonumber
  && - 68 a^3 b + 25 a^4 b - 2 a^5 b -
  74 b^2 - 29 a b^2 + 66 a^2 b^2 - 8 a^3 b^2 - 2 a^4 b^2+ 85 b^3\\ \nonumber
  &&
  - 68 a b^3 - 8 a^2 b^3 + 4 a^3 b^3 - 2 b^4 + 25 a b^4 - 2 a^2 b^4 -
  11 b^5 - 2 a b^5 + 2 b^6.\\
  &&
\end{eqnarray}
The other polynomials are too complicated to be reproduced here. Note that
the Hamiltonian (\ref{E2.1}) with the coupling constants (\ref{E2.2a}, \ref{E2.2b}) commutes with
$\op{S}^2$ and $\op{S}_{23}^2$.
Correspondingly, the indices $i,j$ of the polynomials $q_{ij}$
refer to the quantum numbers $S\equiv i$ and $S_{23}\equiv j$ of the corresponding
competing states, see figure \ref{fige3a}.\\

\begin{center}
\begin{figure}
\begin{center}
  \includegraphics[width=10cm]{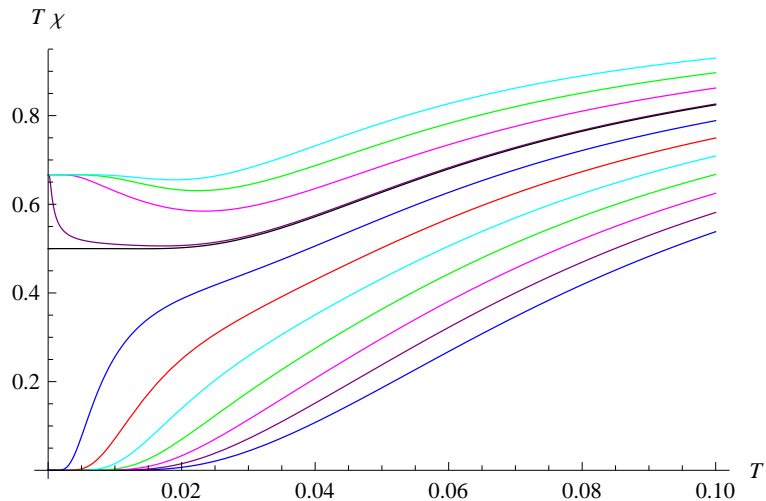}
\end{center}
\caption{\label{fige3c}The product of temperature $T$ and zero-field susceptibility $\chi$
as a function of $T$ for the coupling constants in (\ref{E2.2b}) varying from
$a=0.8$ (lowermost curve) to $a=0.9$ (uppermost curve) and $b=0.4\,a$.
The units are chosen such that $T\chi=\langle M^2 \rangle$ becomes dimensionless.
At the value $a_1=0.869506904299778$ (black curve) the line $b=0.4\,a$
crosses the boundary of the TGS domain and consequently
there is a transition from $0$ to the finite value $1/2$ resp.~$2/3$ of $\lim_{T\to  0}T\chi$  due to the
finite magnetization of the competing state with $S=1$, see figure \ref{fige3a}. }
\end{figure}
\end{center}

\begin{center}
\begin{figure}
\begin{center}
  \includegraphics[width=10cm]{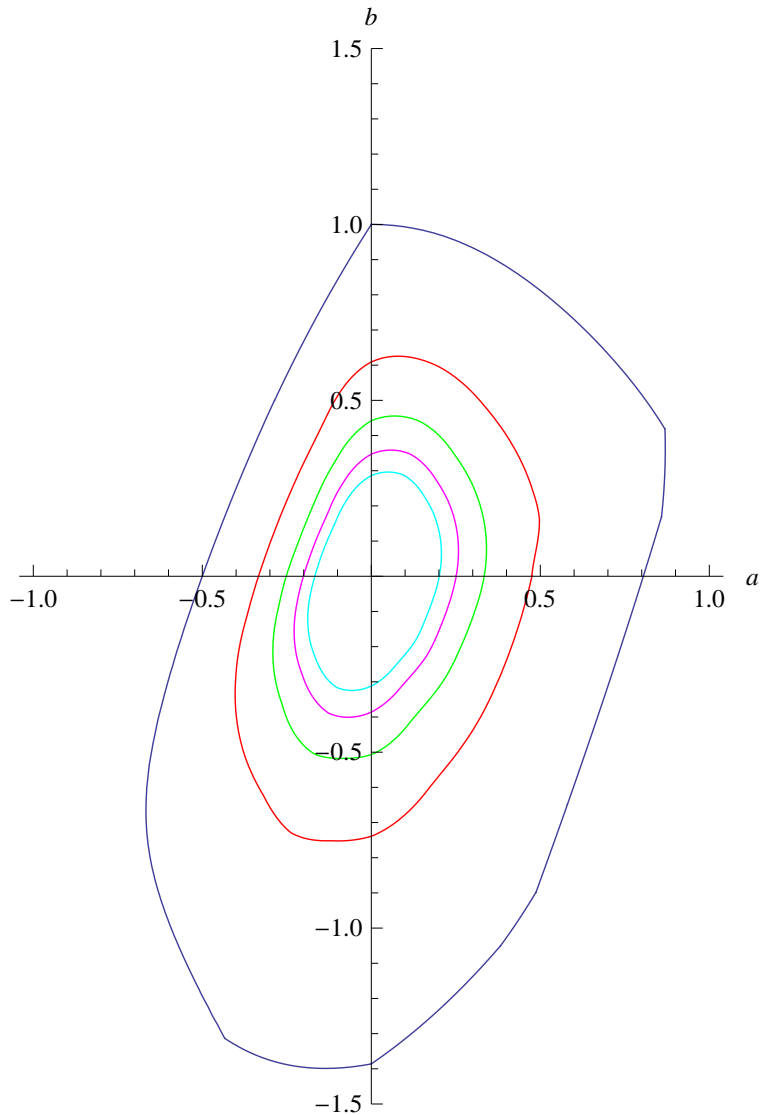}
\end{center}
\caption{\label{fige3b}Representation of the set $\mathcal{T}^s$
of points with coordinates $(a,b)$ such that the singlet product state
$\Phi$ will be a ground state of the Hamiltonian (\ref{E2.1})
corresponding to (\ref{E2.2a},\ref{E2.2b}).
The different curves defining the boundary of $\mathcal{T}^s$ refer
to the values of $s=1,2,3,4,5$ starting from the outermost curve.
The curve for $s=1$ has been calculated analytically, see figure \ref{fige3a},
the other curves for $s=2,3,4,5$ have been determined numerically.}
\end{figure}
\end{center}

In order to illustrate the physical implications of the presence of TGS ground states
for a relatively simple example we have calculated the (dimensionless) zero-field magnetic susceptibility
$\chi =\left.\langle\frac{\partial M}{\partial B}\rangle\right|_{B=0}=\frac{1}{T}\langle M^2 \rangle$ as a function
of temperature $T$ for coupling constants along the line $b=0.4\,a$ and $s=1$, see figure \ref{fige3c}.
This line crosses the boundary of the TGS domain $\mathcal{T}^1$ at the value
$a_1=0.869506904299778$. For $a<a_1$ the ground state  $\Phi$ has $S=0$; for
$a>a_1$ the competing ground states have $S=1$. Hence there is a transition from $\chi(T)$
vanishing exponentially  at $T=0$ to divergence of the form $\chi(T)\sim \frac{2}{3}\frac{1}{T}$.
The factor $\frac{2}{3}$ is simply the mean value of $M^2$ for the three ground states with
$S=1$ and $M=-1,0,1$. For $a=a_1$ the factor is $\frac{1}{2}$, corresponding to the mean
value of $M^2$ for the four ground states with
$S=1$ and $M=-1,0,1$ and $S=M=0$.
This transition is qualitatively the same for gapped infinite TGS chains, see section \ref{sec2.4},
although the factor $\frac{2}{3}$ would have to be replaced by the mean value of $M^2$ of
a continuum of competing ground states.\\

In the two-trimer example the sets $\mathcal{T}^s$ are shrinking when $s$ increases. We generally conjecture that
$\mathcal{C}^s \supset \mathcal{C}^{s'}$ for $s<s'$ but could not prove this rigorously. For $b=0$ and $a<0$ we find
numerically  that the critical value $a_{\mbox{\scriptsize crit}}$ which lies at the boundary of $\mathcal{T}^s$
has the form
\begin{equation}\label{E2.3}
a_{\mbox{\scriptsize crit}}=-\frac{1}{s+1} \mbox{ for } s=1,2,3,4,5
\;.
\end{equation}
This can be confirmed semi-analytically to hold for all $s$
by calculating the competing state $\Psi$ which becomes an additional ground state
if $a$ assumes the value $a_{\mbox{\scriptsize crit}}$. We defer this calculation to the Appendix.

\subsection{Chains}
\label{secE3}
Next we consider examples of chains formed of trimers which are coupled in a balanced way,
i.~e.~satisfying (\ref{S6a})-(\ref{S6d}), such that the
singlet product state
\begin{equation}\label{E3.1}
\Phi=\bigotimes_{i=1}^N\,[i0,i1,i2]
\end{equation}
becomes an eigenstate of the corresponding Hamiltonian.
The coupling within the trimers is always chosen as
\begin{equation}\label{E3.2}
J_{i0,i1}=J_{i1,i0}= J_{i0,i2}=J_{i2,i0}= J_{i1,i2}=J_{i2,i1}=
\frac{1}{2} \;.
\end{equation}
The coupling constants between the trimers
$J_{\delta,\epsilon}\equiv J_{i\delta,\,(i+1)\epsilon}$ are chosen
to depend linearly on one or two parameters and we will investigate
the convex domain of these parameters for which $\Phi$ will be a
ground state of the corresponding Hamiltonian. Throughout this
section we adopt
periodic boundary conditions, i.~e.~$N+1\equiv 1$.\\


\subsubsection{Chains of corner sharing tetrahedra}
\label{secE31}

\begin{center}
\begin{figure}
  \includegraphics[width=15cm]{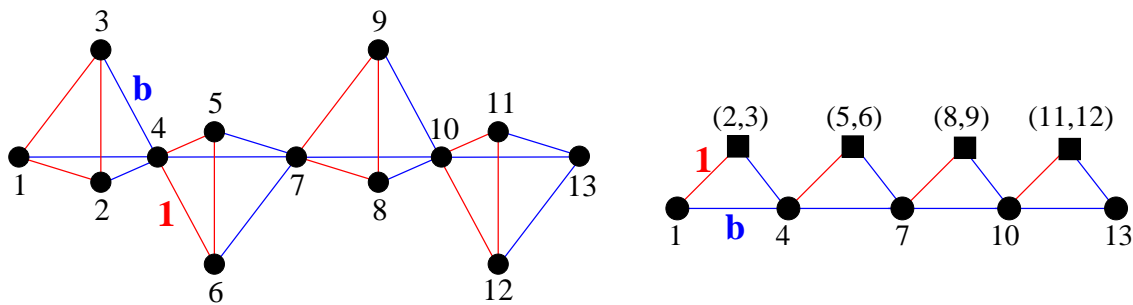}
\caption{\label{fige4}Example of a chain consisting of corner
sharing tetrahedra (left) and an equivalent effective chain of
composite spins (right).}
\end{figure}
\end{center}

The choice of the inter-trimer interaction matrix
\begin{equation}\label{E31.1}
(J_{\delta,\epsilon})=
\frac{1}{2}
\left(
\begin{array}{rrr}
b & 0 & 0\\
b& 0 & 0\\
b&0&0
\end{array}
\right)
\end{equation}
leads to a chain of corner sharing tetrahedra, see figure \ref{fige4}.
Similar systems have been widely considered theoretically as well as
experimentally,
see e.~g.~\cite{tetrahedra}.
However, most of
these studies are focussed on spin-half systems. As it is typical for chains
of corner sharing tetrahedra
the chains considered in this section have
an additional symmetry, namely that certain composite spin squares
(in figure \ref{fige4} these are $\op{S}_{23}^2$, $\op{S}_{56}^2$, etc.~) commute with the Hamiltonian.
Hence we have a simpler sawtooth chain composed of spins and composite
spins, see figure
\ref{fige4}. While in general the composite spins may have spin quantum
number $s_{\mbox{\footnotesize  comp}}=0,\ldots,2s$,  in the singlet product ground state $\Phi$
the composite
spins have the same spin quantum number $s$ as the individual spins.
\\

We have numerically determined the critical values
$b_{\mbox{\footnotesize min}}$ and $b_{\mbox{\footnotesize max}}$
where $\Phi$ ceases to be the unique ground state with energy
$E_0=-\frac{3N}{2}s(s+1)$. These critical values depend on the
number of spins $3N$ and the spin quantum number $s$.  The results
are contained in table \ref{table1}. While the dependence on the
length of the chain is weak, again an increase in the  spin quantum
number leads to a significant smaller parameter region where $\Phi$
is the ground state. From the numerical data we can detect the
competing states which become ground states for
$b<b_{\mbox{\footnotesize min}}$ and $b>b_{\mbox{\footnotesize
max}}$. The competing state for negative, i.e. ferromagnetic, $b$ is
a ferrimagnetic state (the total spin of the chain $S$ is finite but
less than $3Ns$)  and  the composite spins have the spin quantum
number $s-1$. Hence for $s=1$ the spin quantum number of the
composite spins in the competing state is zero. Then the spins along
the base line of the effective sawtooth chain build a simple
ferromagnetic $s=1$ chain which is decoupled from  the composite
spins. The energy of the competing state is $\tilde{E}_0=N(-2 + b)$
and its total spin is $S=N$. As a result there is a prominent
transition for $s=1$ at $b_{\mbox{\footnotesize min}}=-1$. For $s>1$
and feromagnetic $b$ there is no simple competing state, since the
effective chain is a mixed-spin  sawtooth chain \cite{sen04}
(e.g.  a mixed spin-one spin-two
sawtooth chain in case of $s=2$).\\

\begin{center}
\begin{figure}
\includegraphics[width=15cm]{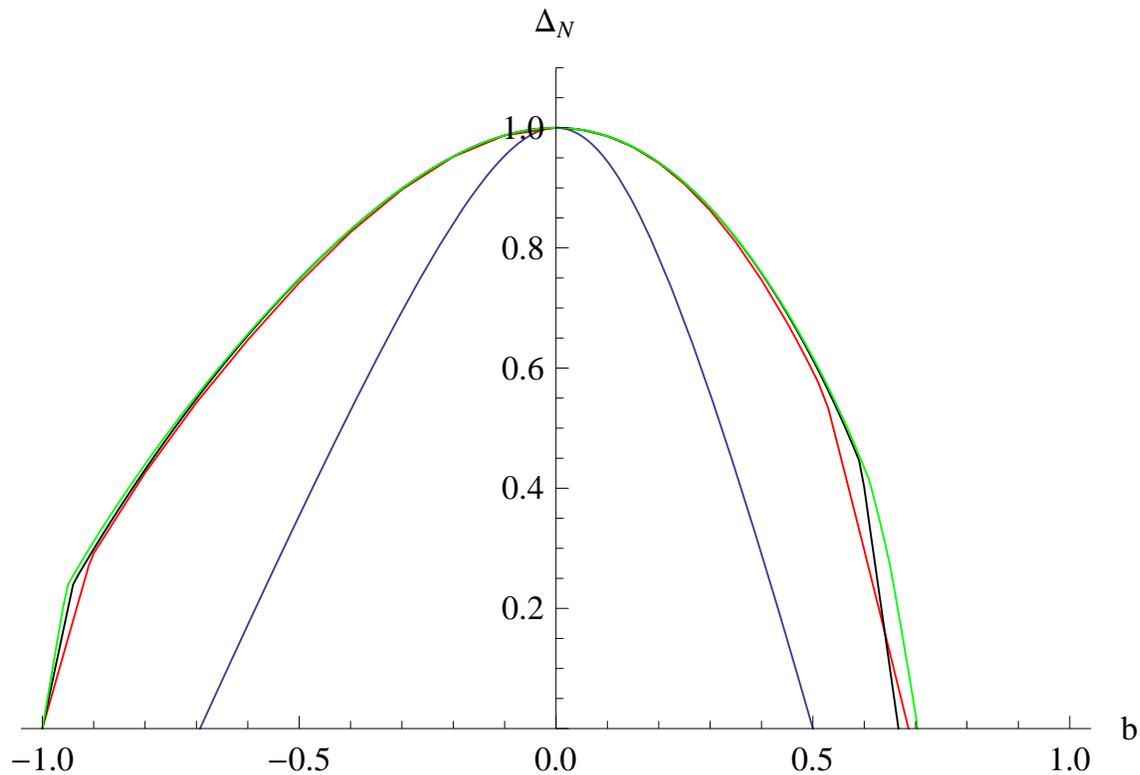}
\caption{\label{fige5a}The numerically calculated energy gap $\Delta_N(b)$ of the chain of $N$ corner sharing tetrahedra
as a function of the coupling constant $b$ for $N=2,3,4,5$ and $s=1$. The lower-most blue curve corresponds to $N=2$;
the next curves correspond to $N=3$ (red), $N=4$ (black) and $N=5$ (green).  According to
corollary $2$ of section \ref{sec2.4} we have the rigorous bound $\Delta_N\ge \Delta_2$ for all $N=3,4,\ldots$, which is confirmed by
these examples. The curves $\Delta_N(b)$ show only small variations for $N=3,4,5$.
They seem to grow monotonically with $N$ except for $b>0.64$ where the $N=3$ and the $N=4$ curves intersect.
The parabolic form of the gap functions in the neighborhood of $b=0$ can be understood by virtue of $1$st and $2$nd order perturbation
theory and considering local $S=1$ excitations for $b=0$ satisfying $\Delta_N(0)=1$.
The $1$st order corrections vanish due to the balanced form of the inter-trimer
Hamiltonian; the $2$nd order contributions have a negative sign since the local excitations can be treated like ground states in the
Hilbert space $\Phi^\perp$. Note that the perturbed local excitations need not be the competing ground states; this explains the
kinks in the gap functions.}
\end{figure}
\end{center}

The competing state for
positive, i.e. antiferromagnetic, $b$ is a state with the total spin of
the chain $S=0$ and  the composite
spins
have the spin quantum number $s-1$. Again for $s=1$ the spin quantum number of the composite
spins in the competing state is zero, and, as a result,
the spins along the base line of the
sawtooth build an antiferromagnetic $s=1$ Haldane chain which is decoupled from  the
composite spins. The energy of the competing state is
$E_0'=N(-2 + b e_0(N))$.
Setting $E_0'=E_0$ yields $b_{\mbox{\footnotesize
max}}=-\frac{1}{e_0(N)}$.
For $N\rightarrow\infty$ the energy of the antiferromagnetic chain is given by $e_0(\infty,1)=-1.40148403897$,
see \cite{White93}. Hence we get for $N\rightarrow\infty$ the critical value
$b_{\mbox{\footnotesize max}}=-1/e_0(\infty,1)=0.713529353$.
This result together with $b_{\mbox{\footnotesize min}}=-1$ indicates that the domain of TGS systems
slightly expands with growing $N$. Hence we expect a non-degenerate TGS domain
even for $N\to\infty$ where $\Phi$ will be a gapped ground state as we will prove in section \ref{sec2.4}.
For $s>1$ the
effective chain is again a mixed-spin  sawtooth
chain and one can find values for $b_{\mbox{\footnotesize max}}$ for short
chains only.
\\
Finally, we have numerically calculated the energy gap $\Delta_N$ of the chain of $N$ corner sharing tetrahedra
as a function of the coupling constant $b$ for $N=2,3,4,5$, see figure \ref{fige5a}. These results
confirm the rigorous bound $\Delta_N\ge \Delta_2$ for all $N=3,4,\ldots$, which will be derived in section
\ref{sec2.4}, corollary \ref{cor2} and shows that the TGS ground state is gapped for trimer chains of this kind.\\

\begin{table}
\caption{\label{table1}Critical values of the coupling constant $b$ corresponding to the chains of figure
\ref{fige4}.}
\begin{center}
\begin{tabular}{|l|l|l|l|}
\hline
$3 N$ & $s$ & $b_{\mbox{\footnotesize min}}$ & $b_{\mbox{\footnotesize max}}$\\
\hline\hline
12 & 1 & -1.0 & 0.667\\
\hline
12 & 2 & -0.57 & 0.434\\
\hline
18 & 1 & -1.0 & 0.696\\
\hline
24 & 1 & -1.0 & 0.706\\
\hline
30 & 1 & -1.0 & 0.710\\
\hline
\end{tabular}
\end{center}
\end{table}

\subsubsection{Spin tubes}
\label{secE32}

\begin{center}
\begin{figure}
\includegraphics[width=17cm]{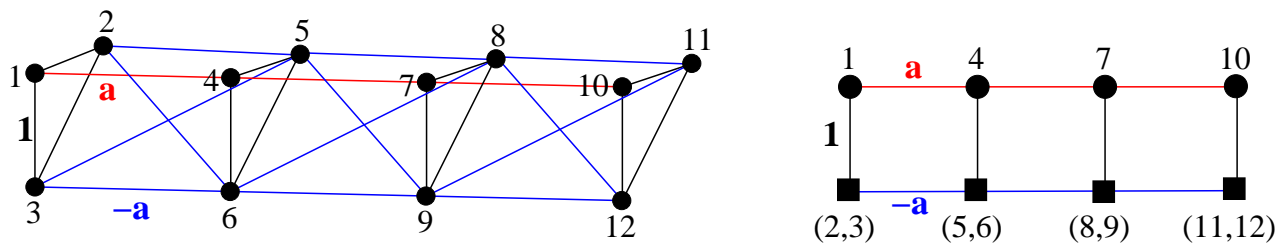}
\caption{\label{fige5}Example of a spin tube I (left) and an equivalent
effective ladder of composite spins (right).}
\end{figure}
\end{center}

Several specific choices of the inter-trimer interaction matrix between the trimers
correspond to so-called three-leg or triangular spin tubes. Such spin tubes
have been widely considered in the literature,
see e.~g.~\cite{tubes}.\\
{\it Spin tube I: }
The choice of the inter-trimer interaction matrix
\begin{equation}\label{E32.1}
(J_{\delta,\epsilon})=
\frac{1}{2}
\left(
\begin{array}{rrr}
a & 0 & 0\\
0& -a & -a\\
0&-a &-a
\end{array}
\right)
\end{equation}
leads to a spin tube as shown figure \ref{fige5}.
 For the special kind of systems (\ref{E32.1})
it turns out that the chains have an additional symmetry, namely that certain composite spin squares
(in figure \ref{fige5} these are $\op{S}_{23}^2$, $\op{S}_{56}^2$, etc.~) commute with the Hamiltonian.
Hence one can consider a simpler ladder model composed of spins and composite spins, see figure \ref{fige5}.
This ladder is frustrated, since the upper and the lower leg exchange bonds have
different sign. In the singlet product state $\Phi$ the composite spins on the
lower leg have the same spin quantum number  $s$ as the individual spins.
Hence we have an effective spin-$s$
ladder, where the antiferromagnetically coupled rungs are in a local
singlet state.  Note that a similar exact singlet product state was also
found  for a spin-half ladder with one ferromagnetic leg and one
antiferromagnetic leg \cite{takahashi}.\\

We have numerically determined
the critical values $a_{\mbox{\scriptsize min}}$ and $a_{\mbox{\scriptsize max}}$
where $\Phi$ ceases to be the unique ground state with energy $E_0=-\frac{3N}{2}s(s+1)$.
These critical values depend on
the number of spins $3N$ and the spin quantum number $s$. The results are contained
in table \ref{table2}. In accordance with our general conjectures we find that the
critical interval $[a_{\mbox{\scriptsize min}},a_{\mbox{\scriptsize max}}]$
shrinks when passing from $s=1$ to $s=2$ and slightly expands from $N=12$ to $N=18$.\\

Based on the numerical data we have analyzed the competing states which become
ground states for $a<a_{\mbox{\scriptsize min}}$ and
$a>a_{\mbox{\scriptsize max}}$. In the competing state for both cases
the composite
spins
have the spin quantum number $s+1$. As a result, the lower leg of the
effective model carrying  larger spins determines the magnetic ordering of
the systems.
While for
positive $a$
within both legs the spin-spin correlations
are ferromagnetic in the competing state,
one has a competing state
with antiferromagnetic  spin-spin correlations within both legs for
negative $a$.
Due to the antiferromagnetic rung coupling
the total spin of the system is $S=0$ for $a<a_{\mbox{\footnotesize min}}$,
whereas
the competing state is
ferrimagnetic with $S=N$ for $a>a_{\mbox{\footnotesize max}}$.
\\

\begin{table}
\caption{\label{table2}Critical values of the coupling constant $a$ corresponding
to the chains of figure \ref{fige5}.}
\begin{center}
\begin{tabular}{|l|l|l|l|}
\hline
$3 N$ & $s$ & $a_{\mbox{\footnotesize min}}$ & $a_{\mbox{\footnotesize max}}$\\
\hline\hline
12 & 1 & -0.319 & 0.418\\
\hline
18 & 1 & -0.330 & 0.419\\
\hline
12 & 2 & -0.210 & 0.270\\
\hline
\end{tabular}
\end{center}
\end{table}

\begin{center}
\begin{figure}
\includegraphics[width=13cm]{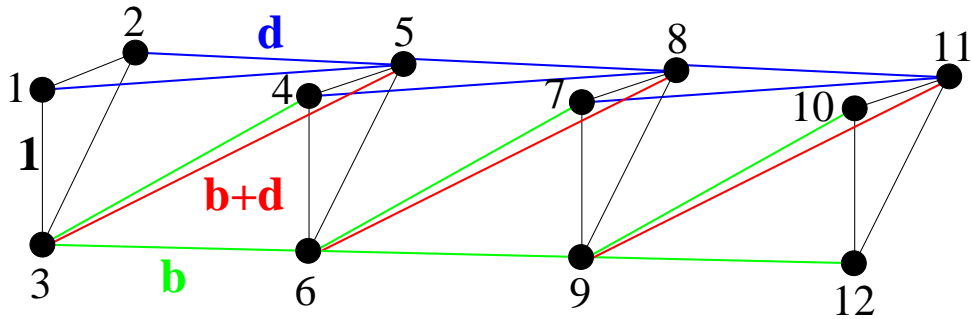}
\caption{\label{fige6}Example of a spin tube II.}
\end{figure}
\end{center}
\begin{center}
\begin{figure}
\includegraphics[width=13cm]{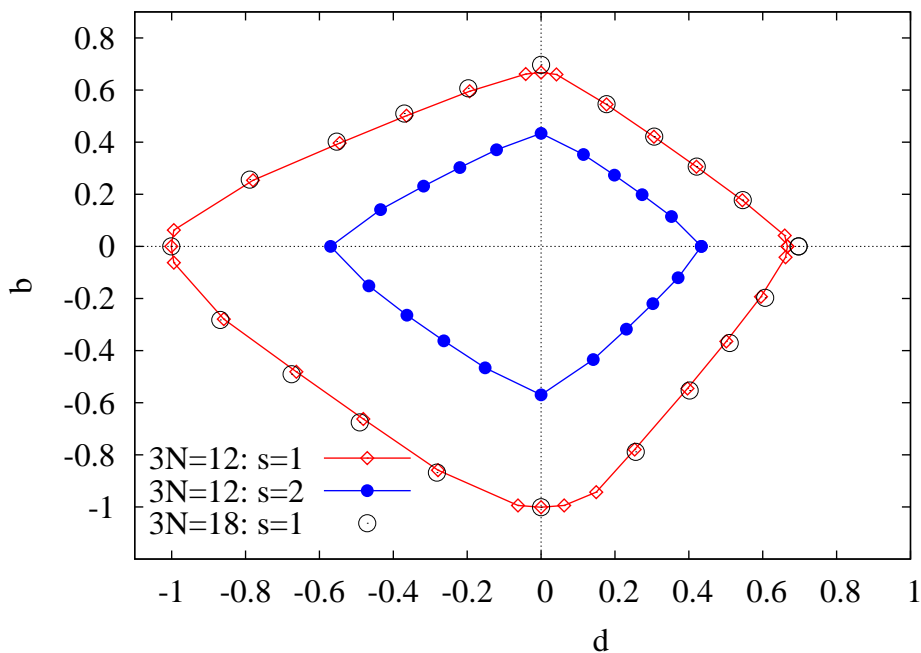}
\caption{\label{fige7}Representation of the set
of points $\mathcal{T}^s$ with coordinates $(b,d)$ such that $\Phi$ will be a ground state of Hamiltonian corresponding to
spin tube II shown in figure \ref{fige6}. }
\end{figure}
\end{center}

{\it Spin tubes II:}
\label{secE33}
The choice of the inter-trimer interaction matrix
\begin{equation}\label{E33.1}
(J_{\delta,\epsilon})=
\frac{1}{2}
\left(
\begin{array}{rrr}
0 & d & 0\\
0& d & 0\\
b&b+d &b
\end{array}
\right)
\end{equation}
leads to another spin tube, see figure \ref{fige6}.  For this
special kind of coupling it turns out that the chains have no
additional symmetry (i.e. no composite spins are conserved), if both
parameters $b$ and $d$ are non-zero. Nevertheless, $\Phi$ will be the unique ground state
for a convex neighborhood of the point $b=d=0$. Our numerical results are
contained in figure \ref{fige7}.

First, we notice that in the limits  $b=0$ or $d=0$ some of the
bonds are missing, and the model can
be transformed to the chain of corner-sharing tetrahedra. As a trivial result the
competing states and the corresponding transition points are the same as discussed in
section \ref{secE31}. Consequently, the competing
state at (and also in the vicinity of) $b=0$, $d=-1$ and also $d=0$, $b=-1$
is ferrimagnetic. In all the other areas we find numerically that the
competing state is a non-trivial singlet state. However, its spin-spin
correlations depend strongly on the position on the transition line shown in
figure \ref{fige7}.

As for the previous examples, it is obvious from figure \ref{fige7}
that there is only a very weak dependence on the size of the system.
Moreover, we observe an inclusion $\mathcal{T}^2\subset\mathcal{T}^1$,
which is compatible with the conjecture that the domain in
the coordinate space $(b,d)$ where $\Phi$ will be a ground state shrinks with
increasing spin quantum number $s$.

\section{Rigorous results}
\label{sec2}

\subsection{Definitions}
\label{sec2.1}
We recall the general definitions given in section \ref{secS}.
Analogous definitions hold for the classical case:
Here the spin observables $\bi{s}_\mu^{\small{cl}}$ are
unit vectors, $H({\mathbb J})^{\small{cl}}$ is the Hamiltonian function,
defined on the $3N$-fold Cartesian product of unit spheres
\begin{equation}\label{2.1.1}
{\mathcal P} \equiv {\begin{array}{c}
{\scriptstyle 3N}\\
{\mbox{\Large\sf X}}
\\^{\scriptstyle \mu=1}\end{array}}
{\mathcal S}_{(\mu)}^2 \;,
\end{equation}
and $\Phi^{\small{cl}}\subset {\mathcal P}$ is the set of all
spin configurations satisfying
\begin{equation}\label{2.1.2}
\bi{s}_{i0} + \bi{s}_{i1} + \bi{s}_{i2}= \bi{0}
\;\mbox{ for all } i=1,\ldots,N
\;.
\end{equation}
Note that  $\Phi^{\small{cl}}$ as well as $\Phi^{\small{s}}$
are invariant under rotations. ${\mathbb J}$ is said to have the
classical TGS property iff the minimum of $H({\mathbb J})$ is assumed for all
$\bi{s}\in \Phi^{\small{cl}}$. In this case we write
${\mathbb J}\in {\mathcal C}_\Phi^{\small{cl}}={\mathcal C}_\Phi^{\infty}$.


\subsection{Necessary conditions for TGS systems}
\label{sec2.2}
Whereas a complete characterization of  ${\mathcal C}_\Phi^s$ seems to be
possible only for small $N$ and $s$ or for the classical case $s=\infty$,
one can prove a number of partial results, either necessary or sufficient
conditions for ${\mathbb J}\in {\mathcal C}_\Phi^s$. \\

We have already mentioned the following result which gives a necessary condition for
${\mathbb J}\in {\mathcal C}_\Phi^s$:
\begin{theorem}
\label{T1}
$\Phi$ is an eigenstate of $\op{H}({\mathbb J})$ iff
\begin{eqnarray}\label{2.2.1a}
J_{i0,j0}+ J_{i1,j1}&=&J_{i0,j1}+J_{i1,j0}\\ \label{2.2.1b}
J_{i0,j0}+ J_{i1,j2}&=&J_{i0,j2}+J_{i1,j0}\\ \label{2.2.1c}
J_{i0,j0}+ J_{i2,j1}&=&J_{i0,j1}+J_{i2,j0}\\ \label{2.2.1d}
J_{i0,j0}+ J_{i2,j2}&=&J_{i0,j2}+J_{i2,j0}
\end{eqnarray}
for all $i<j=2,\ldots,N$. Moreover, let ${\mathcal D} {\mathcal  S}_0$
denote the space of all real $3\times 3$-matrices with vanishing
row and column sums, and $\breve{J}_{ij}$ the $3\times 3$-matrix
with entries $J_{i\epsilon,j\delta},\; \epsilon,\delta=0,1,2$.
Then the above four equations (\ref{2.2.1a}) - (\ref{2.2.1d}) are equivalent
to the statement that $\breve{J}_{ij}$ is orthogonal to the space ${\mathcal D} {\mathcal  S}_0$
w.~r.~t.~the inner product $\langle A,B\rangle=\mbox{Tr }(A^\intercal B)$.
\end{theorem}
{\bf Proof}: The second part of the theorem follows, since the equations
(\ref{2.2.1a})-(\ref{2.2.1d}) say that $\breve{J}_{ij}$ is orthogonal
to the four matrices
\begin{equation}\label{2.2.3}
\left(
\begin{array}{rrr}
1 & -1 & 0\\
-1& 1 & 0\\
0&0&0
\end{array}
\right),
\left(
\begin{array}{rrr}
1 & 0 & -1\\
-1& 0 & 1\\
0&0&0
\end{array}
\right),
\left(
\begin{array}{rrr}
1 & -1 & 0\\
0& 0 & 0\\
-1&1&0
\end{array}
\right),
\left(
\begin{array}{rrr}
1 & 0 & -1\\
0& 0 & 0\\
-1&0&1
\end{array}
\right)
\end{equation}
which span ${\mathcal D} {\mathcal  S}_0$.\\
To prove the first part of the theorem
we rewrite the Hamiltonian (\ref{2.1.1}) in the form
\begin{eqnarray}
\label{2.2.4a}
\op{H}({\mathbb J})
&=&
\sum_{\mu\neq\nu}J_{\mu\nu}
\:\op{\bi{s}}_\mu\cdot \op{\bi{s}}_\mu\\ \nonumber
&=&
\sum_{i\neq j}\sum_{\epsilon,\delta=0}^2 J_{i\epsilon,j \delta}
\:\op{\bi{s}}_{i\epsilon}\cdot \op{\bi{s}}_{j\delta}\\ \label{2.2.4b}
&+& 2 \sum_{i=1}^N \left(
J_{i0,i1} \op{\bi{s}}_{i0}\cdot \op{\bi{s}}_{i1} +
J_{i0,i2} \op{\bi{s}}_{i0}\cdot \op{\bi{s}}_{i2} +
J_{i1,i2} \op{\bi{s}}_{i1}\cdot \op{\bi{s}}_{i2}
\right)  \\     \label{2.2.4c}
&\equiv &
\sum_{i<j}\overline{\op{H}}_{ij}
\;,
\end{eqnarray}
where the distribution of the terms of the second sum in
(\ref{2.2.4b}) to the terms $\overline{\op{H}}_{ij}$ is arbitrary. We have
$\overline{\op{H}}_{ij}=\op{H}_{ij}\otimes \op{\Eins}^{\,(ij)}$ such that  $\op{H}_{ij}$
acts on ${\mathcal H}_{ij}={\mathcal H}_i\otimes{\mathcal   H}_j$ and $\op{\Eins}^{\,(ij)}$
on the remaining factors. Recall that the trimerized state has the form
\begin{equation}\label{2.2.5}
\Phi = \bigotimes_{i=1}^N  [i0,i1,i2]
\;,
\end{equation}
where $[i0,i1,i2]$ denotes the AF trimer ground state in
${\mathcal H}_i={\mathcal  H}_{i0}\otimes {\mathcal  H}_{i1}\otimes {\mathcal  H}_{i2}$.
The following lemma can be proven completely analogous to lemma 2 in \cite{Schmidt05}
\begin{lemma}
\label{L2}
$\Phi$ is an eigenstate of $\op{H}({\mathbb J})$ iff
$[i0,i1,i2]\otimes [j0,j1,j2]$ is an eigenstate of $\op{H}_{ij}$ for all $i<j=2,\ldots N$.
\end{lemma}

In view of this lemma we only need to consider the case of
$N=2$ trimers with indices $i<j$ in the remaining part of the proof.
We set $\phi=[i0,i1,i2]$ and rewrite the indices according to
\begin{equation}\label{2.2.6}
 (i0)\equiv 1,\:  (i1)\equiv 2,\: (i2)\equiv 3,\\
 (j0)\equiv 4,\: (j1)\equiv 5,\: (j2)\equiv 6
 \;.
\end{equation}
Since all summands in
\begin{equation}\label{2.2.7}
0=\langle\phi|S_{123}^2\,\phi\rangle=\sum_{i=1}^3
\langle\phi|(\op{\bi{s}}_1^{(i)}+\op{\bi{s}}_2^{(i)}+\op{\bi{s}}_3^{(i)})^2\,\phi\rangle
\end{equation}
are non-negative, we conclude
$||(\op{\bi{s}}_1^{(i)}+\op{\bi{s}}_2^{(i)}+\op{\bi{s}}_3^{(i)})\,\phi||^2=0$,
i.~e.~$(\op{\bi{s}}_1^{(i)}+\op{\bi{s}}_2^{(i)}+\op{\bi{s}}_3^{(i)})\,\phi=0$ for $i=1,2,3$.
Further, $(\op{\bi{s}}_1+\op{\bi{s}}_2+\op{\bi{s}}_3)\cdot\op{\bi{s}}_\nu(\phi\otimes\phi)=0$
for $\nu=4,5,6$. Hence, for arbitrary $\Psi\in\mathcal{H}_{ij}$, the matrix $D$
with entries
\begin{equation}\label{2.2.8}
D_{\mu\nu}=\langle \Psi|\op{\bi{s}}_\mu\cdot\op{\bi{s}}_{\nu+3}(\phi\otimes\phi)\rangle
\quad \mu,\nu=1,2,3,
\end{equation}
has vanishing row and column sums, i.~e.~$D\in\mathcal{D}\mathcal{S}_0$.
$\phi\otimes\phi$ is an eigenstate of $\op{H}({\mathbb J})$ iff $\breve{\op{H}}(\phi\otimes \phi)=0$
with $\breve{\op{H}}=\sum_{\mu,\nu=1}^3\,\breve{J}_{\mu\nu} \op{\bi{s}}_\mu\cdot\op{\bi{s}}_{\nu+3}$.
This  in turn is equivalent to
$\langle\Psi|\breve{\op{H}}(\phi\otimes\phi)\rangle=0$ for all $\Psi\in\mathcal{H}_{ij}$, or
$\langle \breve{J},D\rangle=\sum_{\mu,\nu=1}^3\,\breve{J}_{\mu\nu}D_{\mu\nu}=0$,
i.~e.~$\breve{J}$ is orthogonal to all matrices in $\mathcal{D}\mathcal{S}_0$ which
can be written in the form (\ref{2.2.8}).\\
It remains to show that there exist enough
$\Psi\in\mathcal{H}_{ij}$ such that the matrices of the form (\ref{2.2.8}) constitute
a basis of $\mathcal{D}\mathcal{S}_0$. Note that for all $s=1,2,3,\ldots\quad \phi$
has a non-vanishing scalar product with the basis vector $e=|1,-1,0\rangle$.
Choose $\Psi=e\otimes e$ and consider the corresponding matrix (\ref{2.2.8}) $D^{(1)}$
with entries $D^{(1)}_{\mu\nu}=\langle\Psi|\op{\bi{s}}_\mu\cdot\op{\bi{s}}_{\nu+3}(\phi\otimes\phi)\rangle=
\langle e|\op{\bi{s}}_\mu \phi\rangle \langle e|\op{\bi{s}}_\nu \phi\rangle$. First, we conclude that
$\langle e|\op{\bi{s}}_\mu^{(1)} \phi\rangle=\langle e|\op{\bi{s}}_\mu^{(2)} \phi\rangle=0$ since
$\op{\bi{s}}_\mu^{(1)}$ and $\op{\bi{s}}_\mu^{(2)}$ change the spin quantum number $S_{123}^{(3)}$
which is $0$ for $e$ and $\phi$. Second, $\op{\bi{s}}_3^{(3)}e=0$, hence
$D^{(1)}=d_1 \left(
\begin{array}{rrr}
1 & -1 & 0\\
-1& 1 & 0\\
0&0&0
\end{array}
\right),\;d_1\neq 0$. The remaining matrices of the basis in (\ref{2.2.3})
are similarly obtained by choosing
$\Psi=|1,-1,0,1,0,-1\rangle,\;\Psi=|1,0,-1,1,-1,0\rangle,$ and $\Psi=|1,0,-1,1,0,-1\rangle$.
This concludes the proof of theorem \ref{T1}.
\hspace*{\fill}\rule{3mm}{3mm}\\

Since  (\ref{2.2.1a})-(\ref{2.2.1d}) is a system of four linearly independent equations,
the set of all
real, symmetric $3N\times 3N$-matrices satisfying (\ref{2.2.1a})-(\ref{2.2.1d}) and
$J_{\mu\mu}=0$ for all $\mu=1,\ldots,3N$ will be a linear space of dimension
$3N+5 {N\choose 2}=\frac{N}{2}(5N+$1), denoted by $\mathcal{J}_\Phi$.
The set $\mathcal{C}_\Phi$ of TGS systems will form a closed convex cone
embedded in the linear space $\mathcal{J}_\Phi$, see \cite{Schmidt05}.\\
If  $\Phi$ is an eigenstate of $\op{H}(\mathbb{J})$ it is straightforward to calculate
the corresponding eigenvalues, since $\langle \Phi|\op{\bi{s}}_{i\delta}\cdot\op{\bi{s}}_{j\epsilon}|\Phi\rangle=0$ for $i\neq j$:
\begin{cor}
If $\Phi$ is an eigenstate of $\op{H}(\mathbb{J})$ then
\begin{equation}\label{2.2.9}
\op{H}(\mathbb{J})\Phi=-s(s+1)\left(\sum_{i=1}^N J_{i0,i1}+ J_{i0,i2}+J_{i1,i2}\right)\;\Phi
\;.
\end{equation}
\end{cor}

In the classical case we have similar but stronger results: The
conditions (\ref{2.2.1a})-(\ref{2.2.1d}) can be strengthened to a uniform
coupling condition:
\begin{theorem}
\label{T2}
If ${\mathbb J}\in {\mathcal C}_\Phi^{cl}$  then
the coupling constants do not depend on $\delta,\epsilon$, i.~e.
\begin{equation}\label{2.2.10}
J_{i\delta,i\epsilon}\equiv J_i>0
\end{equation}
and
\begin{equation}\label{2.2.11}
J_{i\delta,j\epsilon}\equiv\varepsilon_{ij}
\end{equation}
for all $\delta,\epsilon=0,1,2$ and $i<j=2,\ldots,N$.
\end{theorem}
Consequently, we will denote by  ${\mathcal J}_\Phi^\infty\equiv{\mathcal
J}_\Phi^{cl}$ the linear space of all real, symmetric,
$3N\times 3N$-matrices ${\mathbb J}$ with vanishing diagonals and satisfying
(\ref{2.2.10}) (except $J_i>0$) and (\ref{2.2.11}).

\noindent{\bf Proof} of theorem \ref{T2}: The ground states of classical Heisenberg systems satisfy
\begin{equation}\label{2.2.12}
\sum_\nu J_{\mu\nu}\bi{s}_\nu=\kappa_\mu \bi{s}_\mu,\;\mu=1,\ldots,3N
\;,
\end{equation}
see eq.~(16) in \cite{Schmidt03}. This equation results from the condition that the energy
$H=\sum_{\mu\nu} J_{\mu\nu}\bi{s}_\mu\cdot\bi{s}_\nu$ assumes a minimum,
subject to the constraints $\bi{s}_\mu\cdot\bi{s}_\mu=1$. Here the $\kappa_\mu,\;\mu=1,\ldots,3N$
appear as the Lagrange parameters corresponding to these constraints. We choose $\mu=(i,0)$ and rewrite
(\ref{2.2.12}) in the form
\begin{equation}\label{2.2.13}
\sum_{j(\neq i),\epsilon} J_{i0,j\epsilon}\bi{s}_{j\epsilon}
+J_{i0,i1}\bi{s}_{i1}+J_{i0,i2}\bi{s}_{i2}
=\kappa_{i0} \bi{s}_{i0}
\;.
\end{equation}
It is clear, by definition of classical TGS systems, that the contributions
from different trimers with index $j\neq i$ in (\ref{2.2.13}) can be rotated independently.
These rotated contributions cannot be compensated by variations of $\kappa_{i0}$ unless
$\sum_{\epsilon} J_{i0,j\epsilon}\bi{s}_{j\epsilon}$ vanishes for all $j\neq i$.
Choosing $s_{j0}^{(1)}=-s_{j1}^{(1)}=\frac{1}{2}$ and $s_{j2}^{(1)}=0$ yields
$J_{i0,j1}=J_{i0,j2}$. Similar arguments apply to the other equations which say
that the coupling between different trimers must be uniform.\\
To prove uniform coupling within the trimers we reconsider (\ref{2.2.13})
in the form $J_{i0,i1}\bi{s}_{i1}+J_{i0,i2}\bi{s}_{i2}=\kappa_{i0} \bi{s}_{i0}$.
The special choice $s_{i1}^{(1)}=-s_{i2}^{(1)}=\frac{1}{2}$ and $s_{i0}^{(1)}=0$ again yields
$J_{i0,i1}=J_{i0,i2}$ and analogously for the other equations.\\
The previous considerations show that for classical trimerized ground states the Hamiltonian
assumes the value $E=-3 \sum_{i=1}^N J_i$. If one of the $J_i$ would be negative, say $J_1<0$,
one could lower the energy by choosing $\bi{s}_{10}=\bi{s}_{11}=\bi{s}_{12}$. Hence all $J_i\ge 0$
and the proof is complete. \hspace*{\fill}\rule{3mm}{3mm}\\

\subsection{Systems close to unconnected trimers}
\label{sec2.3}

Let $\stackrel{\circ}{\mathbb J}$ denote the matrix of an
{\em unconnected  TGS system}, i.~e.~\\
$(\stackrel{\circ}{J}_{i0,i1},\stackrel{\circ}{J}_{i0,i2},\stackrel{\circ}{J}_{i1,i2})\in\stackrel{\circ}{{\mathcal C}_{\Phi_i}^s}$
for all
$i=1,\ldots,N$ where $\Phi_i$ denote the local trimerized ground states.
All other matrix elements $\stackrel{\circ}{J}_{i\delta,j\epsilon}$
with $i\neq j$ vanish.
Of course, $\stackrel{\circ}{\mathbb J}\in\stackrel{\circ}{{\mathcal C}_\Phi^s}$ and hence
the next lowest energy eigenvalue $E_1$ satisfies
\begin{equation}\label{2.3.0}
E_1=E_0+\varepsilon\equiv\langle\Phi|H(\stackrel{\circ}{\mathbb J})|\Phi\rangle+\varepsilon,\;\varepsilon>0
\;.
\end{equation}
By continuity arguments,
a small neighborhood of $\stackrel{\circ}{\mathbb J}$ still consists
of TGS systems. We want to derive a more quantitative result and consider
an inter-trimer $3N\times 3N$ symmetric coupling matrix
$\Delta\neq 0$ which has to be ``small" in a certain sense.
As a measure of ``smallness" of $\Delta$ we will use $|\delta_{\mbox{\scriptsize min}}|$
where $\delta_{\mbox{\scriptsize min}}$ denotes the lowest eigenvalue of the
matrix $\Delta$.
Note that $\mbox{Tr }\Delta=0$, hence $\delta_{\mbox{\scriptsize min}}<0$ and
the highest eigenvalue $\delta_{\mbox{\scriptsize max}}$ of $\Delta$ satisfies
$0<\delta_{\mbox{\scriptsize max}}\le (3N-1)|\delta_{\mbox{\scriptsize min}}|$.
It is clear that the size of the neighborhood of $\stackrel{\circ}{\mathbb J}$
depends on the energy gap of $H(\stackrel{\circ}{\mathbb J})$
which explains the $\varepsilon$ in the numerator of (\ref{2.3.1}):
\begin{prop}
\label{P3} Let  $\stackrel{\circ}{\mathbb J}$ be an unconnected  TGS
system and  ${\mathbb J}=\stackrel{\circ}{\mathbb J}+\Delta,\;
\Delta\in{\mathcal J}_\Phi$ such that
\begin{equation}\label{2.3.1}
|\delta_{\mbox{\scriptsize min}}| \le \frac{\varepsilon}{ 3N s(s+1)} \;,
\end{equation}
where $\delta_{\mbox{\scriptsize min}}$ denotes the lowest eigenvalue of $\Delta$.
Then ${\mathbb J}\in{\mathcal C}_\Phi^s$.
\end{prop}
Although the proof of proposition \ref{P3} is largely analogous to that of
proposition 3 in \cite{Schmidt05}, we will give it here for sake of convenience.
The $s$-dependence
of the bound in (\ref{2.3.1}) supports the conjecture that the
cones ${\mathcal C}_\Phi^s$ shrink with increasing $s$. \\
\noindent{\bf Proof} of proposition \ref{P3}:
Let $\Psi$ be any normalized state satisfying $\Psi\perp\Phi$. It follows that
\begin{equation}\label{2.3.2}
\langle\Psi|H(\stackrel{\circ}{\mathbb J})|\Psi\rangle \ge E_1
\;.
\end{equation}
Further,
\begin{eqnarray}\label{2.3.3a}
\langle\Psi|H(\Delta)|\Psi\rangle
&=&
\sum_{\mu,\nu}\Delta_{\mu,\nu}\;\langle\Psi|\op{\textbf{s}}_\mu\cdot\op{\textbf{s}}_\nu|\Psi\rangle\equiv \mbox{ Tr }\Delta S\\
&\ge& \delta_{\mbox{\scriptsize min}}\mbox{ Tr } S=\delta_{\mbox{\scriptsize min}}\sum_{\mu=1}^{3N}
\langle\Psi|\op{\textbf{s}}_\mu^2|\Psi\rangle=3N\delta_{\mbox{\scriptsize min}}s(s+1)\\
&\ge&
-\varepsilon
\;,
\end{eqnarray}
where the last inequality follows from (\ref{2.3.1}) and
$\delta_{\mbox{\scriptsize min}}<0$.
It follows that
\begin{eqnarray}\label{2.3.4a}
\langle\Psi|H({\mathbb J})|\Psi\rangle
&=&
\langle\Psi|H(\stackrel{\circ}{\mathbb J})|\Psi\rangle
+
\langle\Psi|H(\Delta)|\Psi\rangle\\
&\ge&
E_1-\varepsilon=E_0
\;,
\end{eqnarray}
hence $\Phi$ will be a ground state of $H({\mathbb J})$.
This concludes the proof of proposition \ref{P3}.
\hspace*{\fill}\rule{3mm}{3mm}\\

An important special case of proposition \ref{P3} is the case of an {\em unconnected homogeneous TGS system},
i.~e.~$\stackrel{\circ}{J}_{i0,i1}=\stackrel{\circ}{J}_{i0,i2}=\stackrel{\circ}{J}_{i1,i2}\equiv \lambda_i>0$ for all
$i=1,\ldots,N$. In this case
\begin{equation}\label{2.3.5}
\varepsilon=2\,\lambda\equiv 2\min\{\lambda_i|i=1,\ldots,N\}
\;.
\end{equation}

\begin{center}
\begin{figure}
  \includegraphics[width=10cm]{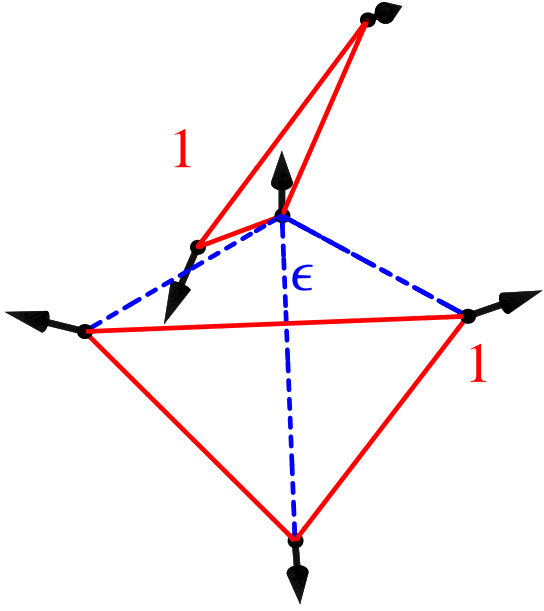}
\caption{\label{fig22}This system cannot be a TGS system for
fixed $\epsilon>0$ and arbitrary $s$, since its classical limit is not TGS.
The classical ground state corresponding to an energy
$E_0=-3-\frac{\epsilon^2}{9}$ is indicated
by small arrows.}
\end{figure}
\end{center}

One of the simplest potential TGS  systems ${\mathbb J}(\epsilon)$,
see figure \ref{fig22}, shows an interesting
effect: For given $s$ and sufficiently small $\epsilon$ it is a
TGS system by virtue of proposition \ref{P3}.
But if $\epsilon>0$
is fixed and $s$ increases, it eventually looses the TGS property.
Otherwise we would get a contradiction since
${\mathbb J}(\epsilon)\notin {\cal C}_\Phi^{cl}$  by theorem
\ref{T2} and the (normalized) ground state energy must converge
for $s\rightarrow\infty$ towards its classical value as a
consequence of the Berezin/Lieb inequality
\cite{Berezin75}
\begin{equation}\label{2.3.10}
(s+1)^2 E_{\mbox{\scriptsize min}}^{\mbox{\scriptsize cl}}
\le
E_{\mbox{\scriptsize min}}
\le
s^2 E_{\mbox{\scriptsize min}}^{\mbox{\scriptsize cl}}
\;.
\end{equation}


\subsection{The TGS chain}
\label{sec2.4}
A TGS chain consists of $N$ copies of trimers such that the intra-trimer and inter-trimer coupling is invariant
under one-dimensional translations and $\Phi$ is a ground state with energy $\tilde{E}_0(N)$. A first question
is whether for a given coupling the system remains a TGS chain for all $N\in\mathbb{N}, N\ge 2$. If this is the case, one
may ask if the difference $\Delta_N$ between the next-lowest eigenvalue $\tilde{E}_1(N)$ and $\tilde{E}_0(N)$
has a positive lower bound independent of $N$. In this case one says that $\Phi$ is a {\it gapped} ground state.
Here we ignore further questions concerning the limit of $\Delta_N$ for $N\to\infty$ and confine ourselves to the
existence of a gap for TGS chains.\\
We will prove our result in a slightly more general context.
Correspondingly, some of the general definitions of the paper are abolished in the following theorem.
We consider a Hamiltonian $H=\sum_{i=1}^N H_i$ where the $H_i$ live in
Hilbert spaces $\mathcal{H}_i\otimes\mathcal{H}_{i+1}$
and the total Hamiltonian  $H$ in $\mathcal{H}=\bigotimes_{i=1}^N\,\mathcal{H}_i$.
All Hilbert spaces $\mathcal{H}_i$ are copies of one finite-dimensional Hilbert space.
These and the following definitions are to be understood in the sense of cyclic boundary conditions $N+1\equiv N$.
Moreover, if $T$ is the unitary translation operator in $\mathcal{H}$ shifting the tensor factors cyclically
and hence satisfying $T^N=\Eins$, we will assume $T\,H_i\,T^\ast=H_{i+1}$ and hence $[T,H]=0$.
\begin{theorem}
\label{T4}
Let $\Phi_i\in\mathcal{H}_i$ be normalized and $\Phi_i\otimes\Phi_{i+1}$ be the unique ground state
of $H_i$ with eigenvalue $E_0$ and the next-lowest eigenvalue being $E_1=E_0+\delta,\;\delta>0$.
Then $\Phi=\bigotimes_{i=1}^N \Phi_i$ will be the unique
ground state of $H$ with eigenvalue $\tilde{E}_0=N\,E_0$ and the next-lowest
eigenvalue satisfies $\tilde{E}_1\ge \tilde{E}_0+2\delta$.
\end{theorem}
{\bf Proof}:
The first claim follows immediately by
\begin{eqnarray}\nonumber
H\Phi&=& \sum_{i=1}^N H_i\Phi =
\sum_{i=1}^N \Phi_1\otimes\cdots\otimes H_i(\Phi_i\otimes \Phi_{i+1})\otimes\cdots \Phi_N\\
&=&\sum_{i=1}^N E_0 \Phi=N E_0\Phi=\tilde{E}_0\Phi
\end{eqnarray}
and $N E_0$ being an obvious lower bound of $H$.\\
Let
$\Psi\in\mathcal{H}$ be the eigenvector of $H$ belonging to the
next-lowest eigenvalues $\tilde{E}_1\ge\tilde{E}_0$. We may assume that
\begin{equation}\label{Psi}
\Psi\perp\Phi \mbox{   and    } T\Psi=e^{i\alpha}\Psi,\;\alpha=2\pi k/N,\, k\in\mathbb{Z}\;.
\end{equation}

Our aim is to show $\tilde{E}_1\ge \tilde{E}_0+2\delta$.
Let $|{\mu}\rangle,\mu=0,1,2,\ldots$ denote the eigenbasis of $H_i$ in $\mathcal{H}_i\otimes\mathcal{H}_{i+1}$
such that $|{0}\rangle = \Phi_i\otimes\Phi_{i+1}$. $|{\mu,K}\rangle$ denotes a corresponding product basis
in $\mathcal{H}$, where $K$ stands for some multi-index of quantum numbers.
Moreover, we consider the reduced density operator $W_\Psi^i$ in $\mathcal{H}_i\otimes\mathcal{H}_{i+1}$
defined by the partial trace
\begin{equation}
\langle\mu|W_\Psi^i|\nu\rangle = \sum_K \langle\mu,K|\Psi\rangle\langle\Psi|\nu,K\rangle
\;.
\end{equation}

Then we conclude
\begin{eqnarray}
\tilde{E}_1&=&\langle \Psi|H|\Psi\rangle = \sum_{i=1}^N \langle \Psi|H_i|\Psi\rangle\\
&=& \sum_{i=1}^N \mbox{Tr}\left( H_i\,W_\Psi^i\right)\\
&=&\sum_{i,\mu} \mbox{Tr}\left( E_\mu |\mu\rangle\langle\mu|W_\Psi^i\right)\\
&=&\sum_{i,\mu}E_\mu\langle\mu|W_\Psi^i|\mu\rangle\\
&=&\sum_i\left(
E_0 \langle 0|W_\Psi^i|0\rangle+\sum_{\mu=1,2,\ldots}E_\mu\langle\mu|W_\Psi^i|\mu\rangle
\right)\\ \label{GE2}
&\ge&
\sum_i\left(
E_0 \langle 0|W_\Psi^i|0\rangle+(E_0+\delta)\sum_{\mu=1,2,\ldots}\langle\mu|W_\Psi^i|\mu\rangle
\right)
\;.
\end{eqnarray}

\begin{lemma}
\begin{equation}\label{lemma2}
\langle 0|W_\Psi^i|0\rangle \le 1-\frac{2}{N}\;.
\end{equation}
\end{lemma}
For the proof of the lemma we use an arbitrary orthonormal basis $|n\rangle,\;n=0,1,2,\ldots$ in
$\mathcal{H}_i$ such that $|0\rangle = \Phi_i$  and $T$ operates
as a cyclic shift operator in the corresponding product basis in $\mathcal{H}$.
Hence $\langle 0|W_\Psi^i|0\rangle $ will be rewritten as $\langle 0,0|W_\Psi^i|0,0\rangle $.
In this notation we have $\Phi=|0,0,\ldots,0\rangle$.
Due to translational symmetry the term (\ref{lemma2}) does not depend on $i$,
hence we may take $i=1$ in what follows. We conclude
\begin{equation}
\langle 0,0|W_\Psi^i|0,0\rangle = \sum_K |\langle\Psi|0,0,K\rangle|^2
\end{equation}
and
\begin{eqnarray} \nonumber
1=\mbox{Tr}W_\Psi^i &=&\sum_{n_3,n_4,\ldots} |\langle\Psi|0,0,n_3,n_4,\ldots\rangle|^2
+ \sum_{n_1,n_2,\ldots}|\langle\Psi|n_1,n_2,\ldots\rangle|^2 \\ \label{sum12}
&\equiv& s_0+s_1
\;.
\end{eqnarray}
The first sum $s_0$ in (\ref{sum12}) runs through all sequences $0,0,n_3,n_4,\ldots$
excluding the value $n_3=n_4=\ldots=0$, since $\langle\Psi|\Phi\rangle=0$.
Equivalently, we will say that it runs through all states $\psi=|0,0,n_3,n_4,\ldots\rangle\in\mathcal{B}_0$.
The second sum $s_1$ in (\ref{sum12}) runs through all sequences $n_1,n_2,\ldots$
except those with $n_1=n_2=0$, or,
equivalently, through all states $\psi=|n_1,n_2,\ldots\rangle\in\mathcal{B}_1$.
Thus the total sum in (\ref{sum12}) runs through an orthonormal basis $\mathcal{B}=\mathcal{B}_0\cup \mathcal{B}_1$ of
$\mathcal{H}'\equiv \{\psi\in\mathcal{H}|\langle\psi|\Phi\rangle=0\}$.\\
We consider on $\mathcal{B}$ the equivalence relation
$\psi_1 \sim \psi_2 \Leftrightarrow \psi_1=T^a\,\psi_2,\;a\in\mathbb{Z},$
and denote by $\Lambda=\mathcal{B}/_\sim$ the corresponding set of equivalence classes or ``orbits".
Due to (\ref{Psi}) all states $\psi$ in the same orbit $\lambda$ yield the same value
\begin{equation}
t_\lambda\equiv |\langle\Psi|\psi\rangle|^2=|\langle\Psi|T^a\,\psi\rangle|^2,\;a\in\mathbb{Z}
\;.
\end{equation}
For each orbit $\lambda\in\Lambda$ let $N_\lambda\equiv|\lambda|$ denote its length.
For most orbits we have $N_\lambda=N$, but in general $N_\lambda$ will be a divisor of $N$.
For example, if $N=6$ and $|1,2,3,1,2,3\rangle\in\lambda$ then $N_\lambda=3$.
We define $N_\lambda^{(k)}\equiv|\lambda\cap\mathcal{B}_k|,\;k=0,1,$ and obtain the
following equations:
\begin{eqnarray}
N_\lambda&=&N_\lambda^{(0)}+N_\lambda^{(1)}\;,\\
s_0&=& \sum_{\lambda\in\Lambda}t_\lambda\,N_\lambda^{(0)}\;,\\
s_1&=& \sum_{\lambda\in\Lambda}t_\lambda\,N_\lambda^{(1)}
\;.
\end{eqnarray}
If $N_\lambda=N$ any basis vector $\psi=|n_1,n_2,n_3,n_4,\ldots\rangle\in\lambda$
has exactly $N$ mutually orthogonal translations. Note that at least one $n_j,\;1\le j\le N$ must
be non-zero since $\psi\neq\Phi=|0,0,\ldots,0\rangle$.
Hence at least two translations of $\psi$ belong to $\mathcal{B}_1$, namely those where $j$ is shifted to $1$ or $2$.
It follows that $N_\lambda^{(1)}\ge 2$ and hence $N_\lambda^{(0)}\le N-2$.
Similarly, in the general case of $1<N_\lambda\le N$ we also have
$N_\lambda^{(1)}\ge 2$ and hence $N_\lambda^{(0)}\le N_\lambda-2\le N-2$.
In the case $N_\lambda=1$, that is, $\psi=|n,n,\ldots,n\rangle,\;n>0$
we have $N_\lambda^{(1)}=1$ and  $N_\lambda^{(0)}=0$. This case has to be treated
separately. We write $\lambda\in\Lambda_1$ iff $N_\lambda=1$ and
$\lambda\in\Lambda_>$ iff $N_\lambda>1$ and conclude
\begin{eqnarray}
s_0&=&\sum_{\lambda\in\Lambda_>}t_\lambda\,N_\lambda^{(0)}\le(N-2)\sum_{\lambda\in\Lambda_>}t_\lambda\;,\\
s_1&=&\sum_{\lambda\in\Lambda_1}t_\lambda+\sum_{\lambda\in\Lambda_>}t_\lambda\,N_\lambda^{(1)}\ge 2\sum_{\lambda\in\Lambda_>}t_\lambda
\;,
\end{eqnarray}
which for $\sum_{\lambda\in\Lambda_>}t_\lambda >0$ implies
\begin{equation}\label{GE}
\frac{s_1}{s_0}\ge\frac{2}{N-2}
\;.
\end{equation}
If $\sum_{\lambda\in\Lambda_>}t_\lambda=0$ then $s_0=0$ and (\ref{lemma2}) follows immediately.
From (\ref{GE}) we infer
\begin{equation}
\frac{1}{s_0}=\frac{s_0+s_1}{s_0}=1+\frac{s_1}{s_0}\ge 1+\frac{2}{N-2}=\frac{N}{N-2}
\end{equation}
and
\begin{eqnarray}
s_0&\le& \frac{N-2}{N}=1-\frac{2}{N}\;,\\ \label{s1}
s_1&\ge& \frac{2}{N}
\;,
\end{eqnarray}
which concludes the proof of the lemma.
\hspace*{\fill}\rule{3mm}{3mm}\\

\noindent To complete the proof of theorem \ref{T4} we consider
\begin{equation}
1=\mbox{Tr }W_\Psi^i=\langle 0|W_\Psi^i|0\rangle+\sum_{\mu=1,2,\ldots}\langle\mu|W_\Psi^i|\mu\rangle
=s_0+s_1
\end{equation}
and rewrite (\ref{GE2}) as
\begin{eqnarray}
\tilde{E}_1&\ge& \sum_{i=1}^N \left(
E_0\,s_0+(E_0+\delta)\,s_1
\right)=
\sum_{i=1}^N (E_0+\delta\,s_1)\\
&=& N\,E_0+N\,\delta\,s_1\ge N E_0+N\,\delta\frac{2}{N}=\tilde{E}_0+2\,\delta
\;,
\end{eqnarray}
where we have used (\ref{s1}) which is equivalent to (\ref{lemma2}).
\hspace*{\fill}\rule{3mm}{3mm}\\
The generalization of theorem \ref{T4} ($d=1$) to square ($d=2$) and cubic ($d=3$) lattices is obvious
but will not be considered here. We only note that in this case the energy gap
is bounded from below by $2^d\,\delta$ for $d=1,2,3$.\\

In order to apply theorem \ref{T4} to trimer chains we will take
\begin{eqnarray}\nonumber
H=H^{(N)}=\sum_{i=1}^N H_i&\equiv& \sum_{i=1}^N
\sum_{\delta,\epsilon=0}^2
\left(
\frac{1}{2}\stackrel{\circ}{J}_{\delta\epsilon}\,
\left(
 \op{\bi{s}}_{i,\delta}\cdot\op{\bi{s}}_{i,\epsilon}+
\op{\bi{s}}_{i+1,\delta}\cdot\op{\bi{s}}_{i+1,\epsilon}
\right)\right.\\ \label{chain}
&+& \left.\breve{J}_{\delta\epsilon}\op{\bi{s}}_{i,\delta}\cdot\op{\bi{s}}_{i+1,\epsilon}
\right)\;,
\end{eqnarray}
where the $\breve{J}_{\delta\epsilon}$ satisfy the conditions of theorem \ref{T1}
and the Hilbert spaces $\mathcal{H}_i,\,i=1,\ldots,N$ are chosen appropriately.
Of course, $\Phi_i=[i_0,i_1,i_2]$.\\
The $3\times 3$-matrix $\breve{J}$ contains five independent real numbers
and may thus be considered as a vector of $\mathbb{R}^5$.
We will fix the values of the intra-trimer coupling $\stackrel{\circ}{J}_{\delta\epsilon}$
such that the open convex set
\begin{equation}\label{T}
\mathcal{T}\equiv \{\breve{J}|H_i \mbox{ has the unique ground state } [i_0,i_1,i_2]\}\subset\mathbb{R}^5
\end{equation}
is non-empty. Hence the energy gap of $H_i$, $E_1-E_0=\delta(\breve{J})$
varies over $\mathcal{T}$ but remains positive there. Then theorem \ref{T4}
shows that $H^{(N)}$ remains a TGS chain for all values $\breve{J}\in\mathcal{T}$ and all $N\in\mathbb{N}$.
Moreover, the energy gap  $\Delta_N(\breve{J})=\tilde{E}_1(N)-\tilde{E}_0(N)$ satisfies $\Delta_N(\breve{J})\ge 2\,\delta(\breve{J})$,
hence $\Phi$ is a gapped ground state in this case. Note that, due to the cyclic boundary conditions,
we have $H^{(2)}=2 H_1$, hence $ \Delta_2(\breve{J})=2\delta(\breve{J})$. We summarize:
\begin{cor}\label{cor2}
Let $H^{(N)}$ be the Hamiltonian of a trimer chain according to (\ref{chain}) and the intra-trimer
coupling $\stackrel{\circ}{J}$ be chosen such that (\ref{T}) is non-empty. Then $H^{(N)}$ will be
a TGS chain for all $\breve{J}\in\mathcal{T}$ and all $N=2,3,\ldots$
and its unique ground state $\Phi$ possesses an energy gap satisfying
$\Delta_N(\breve{J})\ge \Delta_2(\breve{J})$.
\end{cor}

\subsection{The classical case}
\label{sec2.5}

In the classical case it is possible to completely characterize
all TGS systems. Recall that $\epsilon_{ij}=\epsilon_{ji}$ denotes
the uniform interaction strength between two trimers and $J_i$ that
within the trimers according to theorem \ref{T2}.
For any ${\mathbb J}\in{\mathcal J}_\Phi^{cl}$ we
define an $N\times N$-matrix ${\mathbb G}({\mathbb J})$ with entries
\begin{eqnarray}\label{2.4.1a}
G_{ii}&=&J_i \mbox{ for all }i=1,\ldots,N\;,\\ \label{2.4.1b}
G_{ij}&=& \epsilon_{ij} \mbox{ for all }i\neq j=1,\ldots,N\;.
\end{eqnarray}
Then we have the following result:
\begin{theorem}
\label{T3}
Let ${\mathbb J}\in{\mathcal J}_\Phi^{cl}$, then
${\mathbb J}\in{\mathcal C}_\Phi^{cl}$  iff ${\mathbb G}({\mathbb J})$
is positive semi-definite.
\end{theorem}
Recall that ${\Bbb G}({\Bbb J})\ge 0$  iff the $N$ principal minors
$\det(G_{ij})_{i,j=1,\ldots,n}\ge 0$ for $n=1,\ldots,N$.
Hence for classical spin systems the TGS property can be checked
by testing $N$ inequalities. \\

This result is also relevant for quantum spin systems, since we have
the following:
\begin{prop}
\label{P6}
 ${\mathcal  C}_\Phi^{cl}\subset {\mathcal  C}_\Phi^s$ for all
 $s=1,2,3,\ldots$.
\end{prop}
Again, the proofs of theorem \ref{T3} and proposition \ref{P6} are analogous to those
given in sections 5.4 and 5.7 of \cite{Schmidt05}.



\section*{Appendix: Another eigenstate for the trimer pair}

We reconsider the system of two trimers in section \ref{secE2} with the special coupling
\begin{eqnarray}\label{A1a}
J_{12}&=&J_{13}=J_{23}=J_{45}=J_{46}=J_{56}=\frac{1}{2},\\ \label{A1b}
J_{14}&=&-J_{26}=-J_{36}=-J_{25}=-J_{35}=\frac{a}{2},\quad a < 0
\;,
\end{eqnarray}
and the remaining coupling constants vanishing.
We want to calculate a competing eigenstate
$\Psi$ which gives a lower energy than the trimerized state $\Phi$ for $a< a_{\mbox{\scriptsize crit}}$. \\

As usual, we denote the composite spin of a subsystem by subscripts,
e.~g.~$\op{S}_{123}^2=(\op{\bi{s}}_1+\op{\bi{s}}_2+\op{\bi{s}}_3)^2$ with eigenvalues  $S_{123}(S_{123}+1)$.
Recall that it is possible to construct orthonormal bases ${\mathcal B}_T$ for the Hilbert space belonging to a spin system
by means of ``coupling schemes" $T$. For example, the coupling scheme $1\to 12 \to 123$ yields the common eigenbase
of the composite spin squares $\op{S}_{12}^2,\op{S}_{123}^2$ and the $3$-component of the total spin $\op{S}_{123}^{(3)}$.
It may happen that some state $\Psi$ belongs to different ${\mathcal B}_T$'s. In this case $\Psi$ will be
a common eigenvector of all composite spin squares corresponding to the different coupling schemes $T$.
For example, the $S=0$ ground state $[1,2,3]$ of a uniform AF trimer belongs to both coupling
schemes $1\to 12 \to 123$ and $1\to 13 \to 123$ and hence has the good quantum numbers $S_{12}=S_{13}=s$.
(Otherwise it cannot couple with, say, $S_3=s$ to give $S_{123}=0$).
Note that a vector $\Psi$ will belong to different ${\mathcal B}_T$'s if it is already uniquely determined by a proper subset
of the quantum numbers corresponding to some coupling scheme $T_1$. If $T_2$ is any other scheme containing the same quantum numbers
we have necessarily $\Psi\in {\mathcal B}_{T_2}$.\\

We will apply these considerations to the coupling scheme $T_1=(2\to 23\to 123, 5\to 56\to 456\to 123456)$ and the
quantum numbers $S_{23}=S_{56}=s+1,\; S_{123}=1,\;S_{123456}=S=0$. Obviously, it follows that $S_{456}$ must have the
value $1$ and the vector $\Psi$ is uniquely determined by these quantum numbers. The same subsystems
$23,56,123,123456$
also occur in the
coupling scheme $T_2=(5\to 56, 2\to 23\to 123 \to 1234 \to 123456)$. Hence $\Psi\in {\mathcal B}_{T_1}\cap {\mathcal B}_{T_2}$
and thus $\Psi$ must have the quantum number $S_{1234}=s+1$. Analogously, $S_{1456}=s+1$.
It follows that $\Psi$ is an eigenstate of the Heisenberg Hamiltonian
\begin{eqnarray}\nonumber
\op{H}_1&=& \left( \op{S}^2-\op{S}_{123}^2-\op{S}_{456}^2\right)\\ \nonumber
&&-\left( \op{S}_{1234}^2-\op{S}_{123}^2-\op{S}_{4}^2\right)
-\left( \op{S}_{1456}^2-\op{S}_{456}^2-\op{S}_{1}^2\right)\\ \label{A2}
&=&2( -\op{\bi{s}}_1\cdot\op{\bi{s}}_4+
\op{\bi{s}}_2\cdot\op{\bi{s}}_5+\op{\bi{s}}_2\cdot\op{\bi{s}}_6+\op{\bi{s}}_3\cdot\op{\bi{s}}_5+\op{\bi{s}}_3\cdot\op{\bi{s}}_6)
\end{eqnarray}
with eigenvalue $E_1=-4(s+1)$. If $\op{H}(a)$ denotes the Hamiltonian according to (\ref{E2.2a}), (\ref{E2.2b}) and $b=0$, we conclude
that
\begin{equation}\label{A3}
\op{H}(a)=\left(
\op{S}_{123}^2-3s(s+1)+\op{S}_{456}^2-3s(s+1)
\right)-a\,\op{H}_1
\;,
\end{equation}
and hence
$\langle \Psi|\op{H}(a)|\Psi\rangle = 4a(s+1)+4-6s(s+1)$. This equals $\langle \Phi|\op{H}(a)|\Phi\rangle = -6s(s+1)$ for
$a=a_{\mbox{\scriptsize crit}}=-\frac{1}{s+1}$ which confirms (\ref{E2.3}).
Note that the argument is not completely
rigorous, since we could not exclude other competing ground states than $\Psi$ for arbitrary $s$.
However, it is in agreement with our numerical data presented above.


\section*{Acknowledgement}
The numerical calculations were performed using J.~Schulenburg's
{\it spinpack}. We thank Klaus B\"arwinkel and Jochen Gemmer for a critical examination of parts of the paper
and stimulating discussions. J.~R.~acknowledges financial support of the DFG (project no.~RI615/16-1).


\end{document}